\documentstyle[12pt,epsfig]{article}
\begin{document}
\title{Aspects of thermal and chemical equilibration
of hadronic matter
\thanks{Work supported by BMBF, GSI Darmstadt and DFG.}}
\author{E. L. Bratkovskaya, W. Cassing, C. Greiner,\\
M. Effenberger, U. Mosel and A. Sibirtsev \\
{\normalsize Institut f\"{u}r Theoretische Physik, Universit\"{a}t Giessen}\\
{\normalsize 35392 Giessen, Germany}}
\date{}
\maketitle
\begin{abstract}
We study thermal and chemical equilibration in 'infinite' hadron matter
as well as in finite size relativistic nucleus-nucleus collisions using
a BUU cascade transport model that contains resonance and string
degrees-of-freedom.  The 'infinite' hadron matter is simulated within a
cubic box with periodic boundary conditions.  The various equilibration
times depend on baryon density and energy density and are much shorter
for particles consisting of light quarks then for particles including
strangeness.  For kaons and antikaons the chemical equilibration time
is found to be larger than $\simeq$ 40 fm/c for all baryon and energy
densities considered.  The inclusion of continuum excitations, i.e.
hadron 'strings', leads to a limiting temperature of $T_s\simeq$ 150
MeV. We, furthermore, study the expansion of a hadronic fireball after
equilibration. The slope parameters of the particles after expansion
increase with their mass; the pions leave the fireball much faster then
nucleons and accelerate subsequently heavier hadrons by rescattering
('pion wind').  If the system before expansion is close to the limiting
temperature $T_s$, the slope parameters for all particles after
expansion practically do not depend on (initial) energy and baryon
density.  Finally, the equilibration in relativistic nucleus-nucleus
collision is considered.  Since the reaction time here is much shorter
than the equilibration time for strangeness, a chemical equilibrium of
strange particles in heavy-ion collisions is not supported by our
transport calculations.  However, the various particle spectra can
approximately be described within the blast model.
\end{abstract}
\vspace{2mm}



\newpage
\section{Introduction}

Nucleus-nucleus collisions at relativistic and ultrarelativistic
energies are studied experimentally and theoretically to obtain
information about the properties of hadrons at high density and/or
temperature as well as  about the phase transition to a new state of
matter, the quark-gluon plasma (QGP).  In the latter deconfined partons
are the essential degrees of freedom that resolve the underlying
structure of hadrons \cite{QM}.  Whereas the early 'big-bang' of the
universe most likely evolved through steps of kinetic and chemical
equilibrium, the laboratory 'tiny bangs' proceed through phase-space
configurations that initially are far from an equilibrium phase and
then evolve by fast expansion. These 'specific initial conditions' --
on the theoretical side -- have lead to a rapid development of
nonequilibrium quantum field theory and nonequilibribrium kinetic
theory \cite{BotMal90,Henning}.  Presently, semiclassical transport
models are widely used as approximate solutions to these theories and
practically are an essential ingredient in the experimental data
analysis. For recent reviews we refer the reader to Refs.
\cite{Ko,Bass,Cass99}.

On the other hand, many observables from strongly interacting systems
are dominated by many-body phase space such that spectra and abundances
look 'thermal'.  It is thus tempting to characterize the experimental
observables by global thermodynamical quantities like 'temperature',
chemical potentials or entropy \cite{BM,Satz,Sollfrank,Spieles,Cleymans}.
We note, that even the use of macroscopic models like hydrodynamics
\cite{Hydro,Rischke} employs as basic assumption the concept of local
thermal and chemical equilibrium. The crucial question, however, how
and on what timescales a global thermodynamic equilibrium can be
achieved, is presently a matter of debate. Thus nonequilibrium
approaches have been used in the past to address the problem of
timescales associated to global or local equilibration
\cite{Rafelski,Cass90,Lang91,Bl93,Brav1,Brav2,Brav3,Solfr99}.  In view
of the increasing 'popularity' of thermodynamic analyses a thorough
microscopic reanalysis of this questions appears necessary especially
for nucleus-nucleus collisions at ultrarelativistic energies that aim
at 'detecting' a phase transition to the QGP.

In this paper we study equilibration phenomena in 'infinite' hadronic
matter using a microscopic transport model that contains both hadron
resonance and string degrees-of-freedom.  With this investigation we
want to provide insight into the reaction dynamics by the use of
cascade-like models and also point out some of their limitations.  The
'infinite' hadronic matter is modelled by initializing the system
solely by nucleonic degrees of freedom through a fixed baryon density
and energy density, while confining it to a cubic box and imposing
periodic boundary conditions during the propagation in time.  We,
furthermore, then study the expansion of the hadronic fireball after
equilibration to investigate the changes in hadron spectra during the
rapid explosion as well as related equilibration phenomena in realistic
nucleus-nucleus collisions for light and heavy systems.

Our paper is organized as follows: In Section 2 we briefly describe the
approach employed in our investigations, specify the initial conditions
for a finite box with periodic boundary conditions, present our
numerical results and extract various (hadronic) equilibration times as
well as thermodynamical properties for different initial conditions.
Section 3 is devoted to the expansion dynamics of the equlibrated
fireball and a discussion of the related physical phenomena. In Section
4 we analyse reactions of colliding finite light and heavy systems and
compare our result to a blast model. Section 5 concludes our study
with a summary.

\section{Equilibration and limiting temperature}

To investigate the equilibration phenomena addressed above we perform
microscopic calculations using the Boltzmann-Uehling-Uhlenbeck (BUU)
model of Refs. \cite{Effe99gam,EffePhD}. This model is based on the
resonance concept of nucleon-nucleon and meson-nucleon interactions at
low invariant energy $\sqrt{s} \ $ \cite{TeisZP97}, adopting all
resonance parameters from the Manley analysis \cite{Manley}.  The high
energy collisions -- above $\sqrt{s}$ = 2.6~GeV for baryon-baryon
collisions and $\sqrt{s}$ = 2.2~GeV for meson-baryon collisions -- are
described by  the LUND string fragmentation model FRITIOF
\cite{FRITIOF}. This aspect is similar to that used in the HSD approach
\cite{Cass99,Ehehalt,Brat98,Geiss} and the UrQMD code \cite{Bass}. For
a detailed description of the underlying model at low energy we refer
the reader to Ref.~\cite{EffePhD}.

For later discussions it is essential to realize that the code respects
detailed balance only for reactions of the type $1 \leftrightarrow 2 +
3$ and approximately for $1 + 2 \leftrightarrow 3 + 4$
\footnote{In the latter case small violations of detailed balance are
due to the treatment of $t$-channel and background contributions.}
where the numbers $1, \ldots, 4$ are any reaction partners. This
implies that in particular at high energies, where the string degrees
of freedom with their decay to many ($> 2$) final particles becomes
important, detailed balance is violated. We will discuss the
consequences of this violation, which is inherent in all such transport
codes, at the appropriate points in the following sections.

\subsection{A box with periodic boundary conditions}

In order to study 'infinite' hadronic matter problems we confine the
particles in a cubic box with periodic boundary conditions for their
propagation similar to a recent box calculation within the UrQMD model
\cite{Brav1}. We specify the initial conditions, i.e. baryon density
$\rho$, strange particle density $\rho_S$ and energy density
$\varepsilon$ as follows: first the initial system is fixed to $N_p=80$
protons and $N_n=80$ neutrons, which are randomly distributed in a
cubic box of volume $V$. The 3-momenta $\vec p_i$ of the nucleons in a
first step are randomly distributed inside a Fermi-sphere of radius
$p_F$ = 0.26~GeV/c (at $\rho_0$) and in a second step  boosted by $\pm
\beta_{cm}$ by a proper Lorentz transformation. Thus the initial baryon
density $\rho$ is fixed as $\rho=A/V$, $A=N_p+N_n$. The strange
particle density is set to zero as in related heavy-ion experiments
while the energy density is defined as $\varepsilon = E/V$, where $E$
is the total energy of all nucleons
\begin{eqnarray}
E=\sum\limits_i^A\sqrt{p_i^2+m_N^2}.
\label{energy_i}
\end{eqnarray}
The  boost velocity $\beta_{cm}$ is related to the initial energy density
$\varepsilon$ (excluding Fermi motion) as
\begin{eqnarray}
\beta_{cm}=\sqrt{1-{\rho^2 m_N^2\over \varepsilon^2} }
\label{betta}\end{eqnarray}
using
\begin{eqnarray}
\varepsilon = \gamma_{cm} \rho m_N \label{vargam}
\end{eqnarray}
with $\gamma_{cm}= 1/\sqrt{1-\beta_{cm}^2}$. Recall that $\rho_0 m_N \simeq
0.15$~GeV/fm$^3$ so that an energy density $\varepsilon \simeq
1.5$~GeV/fm$^3$ at density $\rho_0$ corresponds to $\gamma_{cm}\simeq
10$, i.e. the SPS energy $T_{lab}\simeq 185$~A$\cdot$GeV. We thus start
with a 'true' nonequilibrium situation in order to mimique the  initial
stage in a relativistic heavy-ion collision. The initial phase
represents two interpenetrating, (ideally) infinitely extended fluids
of cold nuclear matter.

We now propagate all particles in the box in the cascade mode (without
mean-field potentials) using periodic boundary conditions, i.e.
particles moving out of the box are reinserted at the opposite side
with the same momentum. The phase-space distribution of particles then
can change due to elastic collisions,  resonance and string production
and their decays to mesons and baryons again. We recall that we include
all baryon resonances up to an invariant mass of 2 GeV and meson
resonances up to the $\phi$-meson. According to the initial conditions
for $\varepsilon$ and $\rho$ the factor $\gamma_{cm}$ in (\ref{vargam})
determines if strings are excited in the very first collisions. This is
the case for $\gamma_{cm} > 1.4$ where the early equlibration stages
are dominated by string formation and decay.

\subsection{Chemical equilibration}

\begin{figure}[t]
\phantom{a}\vspace*{-2cm}
\centerline{\psfig{figure=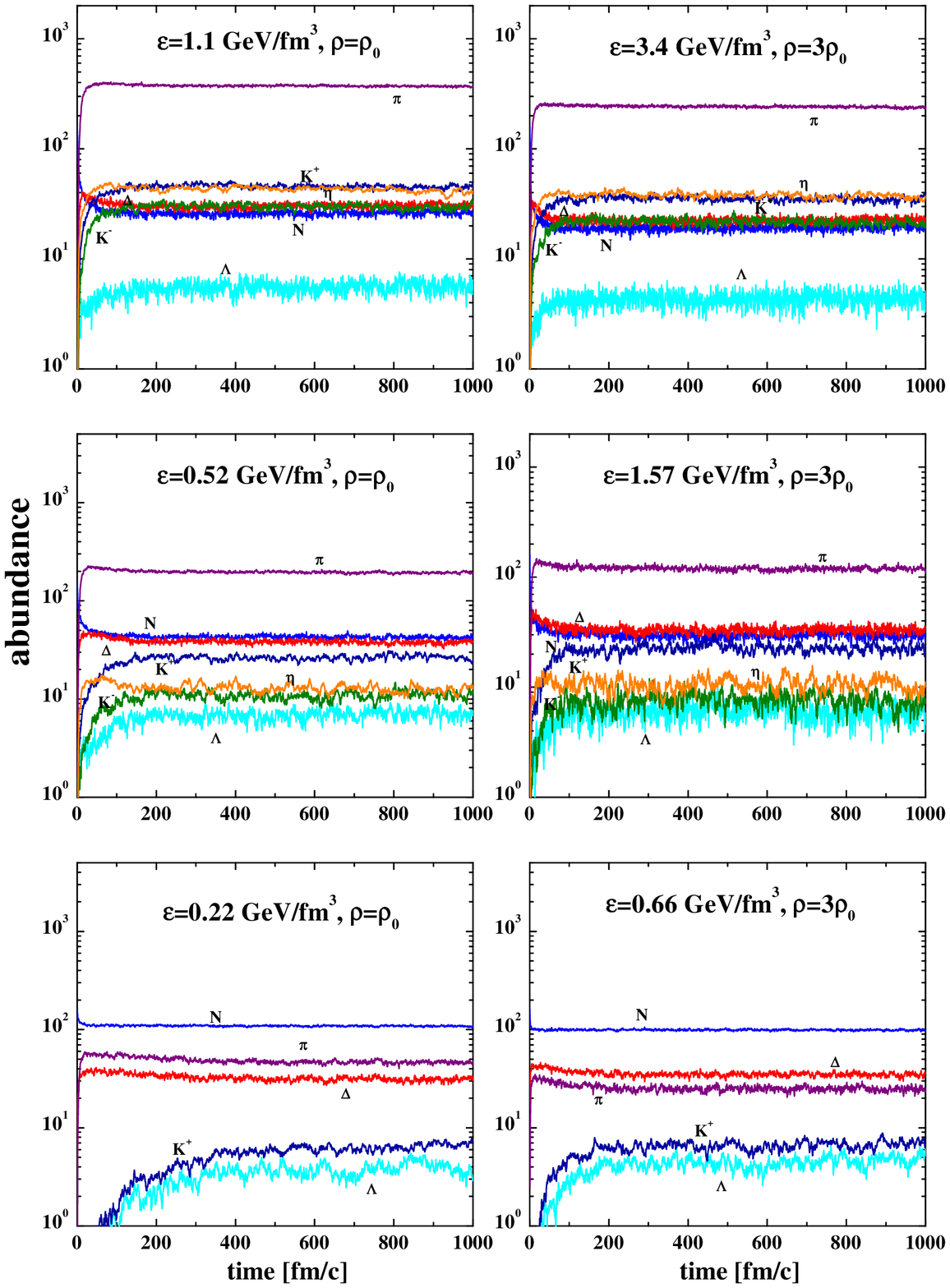,width=16cm}}
\vspace*{-2.5cm}
\caption{Time evolution of the various
particle abundances (nucleons $N$,  $\Delta$,  $\Lambda$, $\pi$,
$\eta$,  $K^+$ and $K^-$ mesons) for density $\rho=\rho_0$ (left
panel) at different energy densities $\varepsilon =1.1, 0.52$ and
0.22 GeV/fm$^3$ and for $\rho=3\rho_0$ (right panel) at
$\varepsilon =3.4, 1.57$ and 0.66 GeV/fm$^3$.}
\label{Fig1}
\end{figure}

\begin{figure}[t]
\phantom{a}\vspace*{-2cm}
\centerline{\psfig{figure=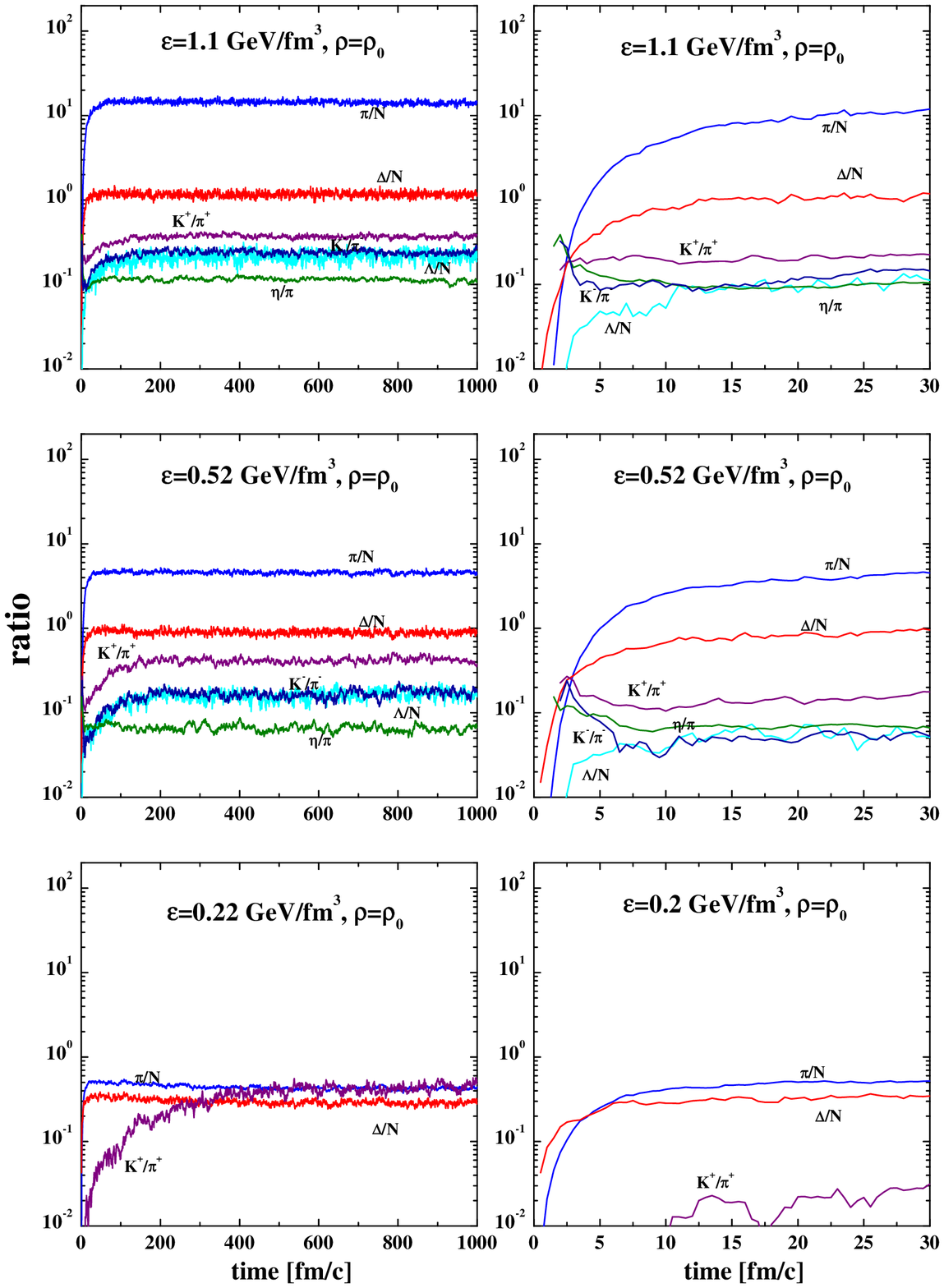,width=16cm}}
\vspace*{-2.5cm}
\caption{Time evolution of particle ratios $\pi/N$, $\Delta/N$,
$\Lambda/N$, $K^+/\pi^+$, $K^-/\pi^-$, $\eta/\pi$ for density
$\rho=\rho_0$ at energy densities $\varepsilon = 1.1, 0.52$ and 0.2
GeV/fm$^3$. The left panel shows the time scale up to 1000 fm/c,
whereas the right panel demonstrates the initial stage up to 30 fm/c.}
\label{Fig2}
\end{figure}

\begin{figure}[t]
\phantom{a}\vspace*{-2cm}
\centerline{\psfig{figure=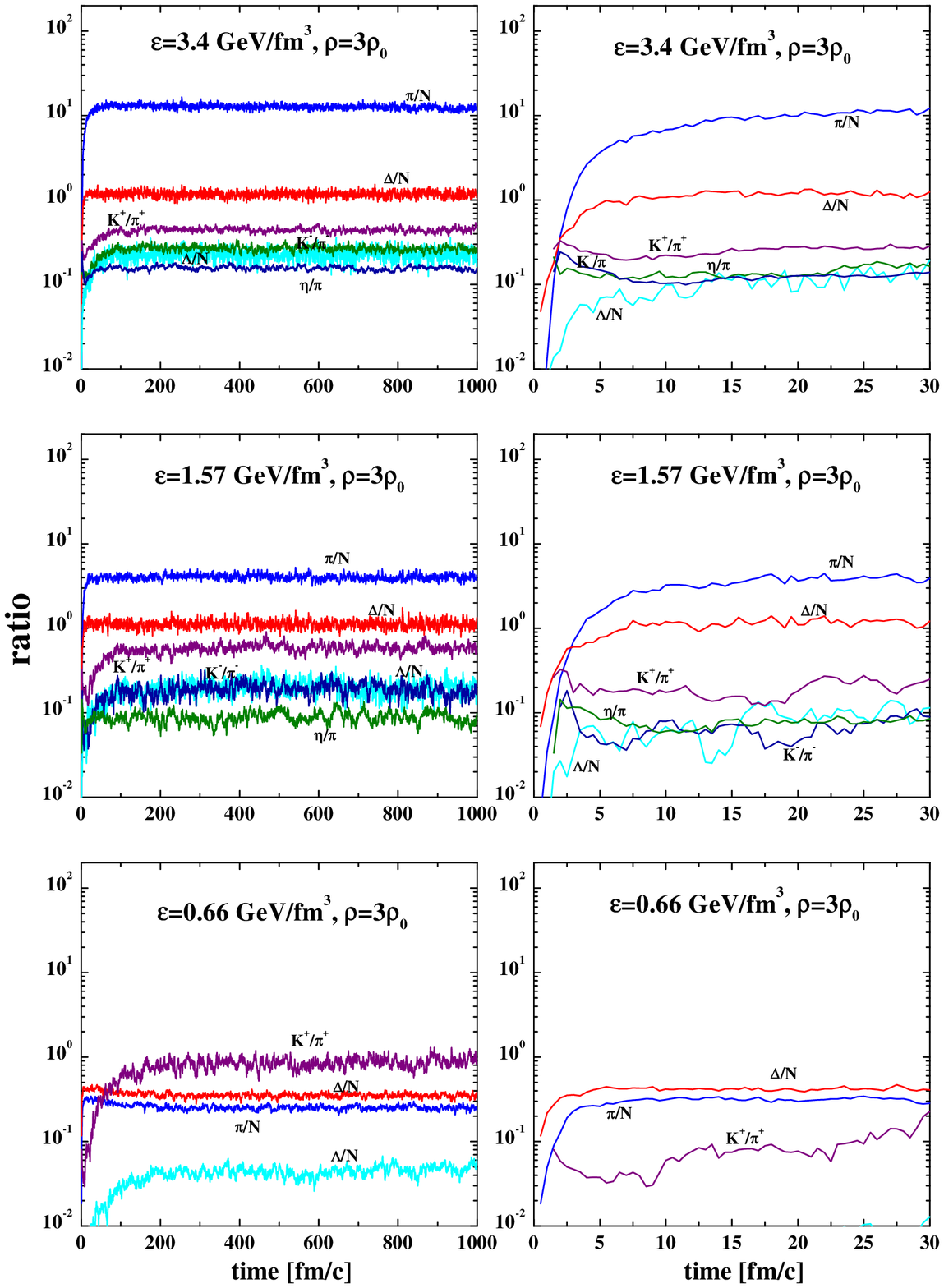,width=16cm}}
\vspace*{-2.5cm}
\caption{Time evolution of particle ratios $\pi/N$, $\Delta/N$,
$\Lambda/N$, $K^+/\pi^+$, $K^-/\pi^-$, $\eta/\pi$ for density $\rho=3\rho_0$
at energy densities $\varepsilon = 3.4, 1.57$ and 0.66 GeV/fm$^3$.
The left panel shows the time scale up to 1000 fm/c, whereas the right
panel demonstrates the initial stage up to 30 fm/c.}
\label{Fig3}
\end{figure}

Figure \ref{Fig1} shows the time evolution of the various particle
abundances (nucleons $N$,  $\Delta$,  $\Lambda$, $\pi$, $\eta$,  $K^+$
and $K^-$ mesons) for density $\rho=\rho_0$ (left panel) at different
energy densities $\varepsilon =1.1, 0.52$ and 0.22 GeV/fm$^3$ and for
$\rho=3\rho_0$ (right panel) at $\varepsilon =3.4, 1.57$ and
0.66~GeV/fm$^3$.  These initial conditions correspond to bombarding
energies $T_{lab}$ per nucleon of roughly 100, 20 and 2 A$\cdot$GeV,
respectively. In Fig.~\ref{Fig1} (as well as in
Figs.~\ref{Fig2},\ref{Fig3}) we count all particles which are
'hadronized', i.e.\ produced by string decay after a  formation time of
$\tau_F=0.8$~fm/c in their rest frame.

After several fm/c the number of nucleons decreases due to inelastic
collisions that produce either baryon resonances or additional mesons.
The number of $\Delta$-resonances grows up to a maximum in a few fm/c,
since a lot of $\Delta$'s are produced in the first $NN$ collisions;
their number subsequently decreases with time due to their decay and
excitation of further resonances or due to reabsorption.  The numbers
of $\pi$'s  and $\eta$'s increase very fast and reach the equilibrium
value within a few fm/c whereas the strange particles ($K^+,
K^-,\Lambda$) require a much longer time for equilibration.

In Fig. \ref{Fig2} we present the time evolution of the particle ratios
$\pi/N$, $\Delta/N$, $\Lambda/N$, $K^+/\pi^+$, $K^-/\pi^-$, $\eta/\pi$
for density $\rho=\rho_0$ at energy densities $\varepsilon = 1.1, 0.52$
and 0.2 GeV/fm$^3$, while Fig.~\ref{Fig3} shows the same particle ratios
for density $\rho=3\rho_0$ at energy densities $\varepsilon = 3.4,
1.57$ and 0.66 GeV/fm$^3$, respectively. The left panels in both plots
correspond to the full time scale as in Fig.~\ref{Fig1} (up to 1000 fm/c),
whereas the right panels present in more detail the initial phase (up
to 30 fm/c).  We use the same scale for the $y$-axis on the right and
left panels, so one can easily see that the $\pi/N$, $\Delta/N$ ratios
reach the equilibrium values very fast especially at low energy
density since the string degrees of freedom here play a minor role and
pion production basically emerges through $\Delta$ resonance decay.

The meson-pion ratios ($K^+/\pi^+$, $K^-/\pi^-$, $\eta/\pi$) at high
energies show a decrease in the first few fm/c and then an increase
again up to the equilibrium values. This is due to the fact that the
bulk of the strange mesons is produced very early (at high energy
density) through string formation and decay whereas most of the pions
appear later, with a delay of several fm/c, as a result of the decay of
heavy vector mesons (e.g. $\rho$ and $\omega$). From the right panels
of Figs.~\ref{Fig2} and \ref{Fig3} one can see that in the initial
stage the particle ratios containing strange to nonstrange particles --
$K^+/\pi^+$, $K^-/\pi^-$, $\Lambda/N$ -- are still far off chemical
equilibrium for all energies and densities and the equilibration takes
up to a few hundred fm/c depending on the energy and baryon density.

For the higher energies the initial particle production proceeds via
the formation and decay of string excitations. This leads in particular
to a very early onset of strange particles (mainly kaons and hyperons)
within the first fm/c either due to the initial strings or due to
secondary or ternary baryon-baryon, meson-baryon and meson-meson
induced string-like interactions (see the right panels of Figs.
\ref{Fig2} and \ref{Fig3}). In Ref. \cite{Geiss} it was shown that
these early secondary and ternary reactions can contribute up to about
50 $\%$ of the total strange particles obtained in a Pb~+~Pb reaction at
CERN SPS energies and thus explain the factor of 2 in the observed
relative strangeness enhancement compared to p+p reactions. This,
however, does not imply that chemical equilibrium for the dominant
strange particles has been achieved in this reaction, as our analysis
clearly shows. In the later stages, when the system has become, more or
less, isotropic in momentum space, strange particles can only be
further produced by low energy hadronic reactions, which, however, have
a considerable threshold and are thus strongly suppressed. This
explains the long chemical equilibration times for the strange
particles first demonstrated by Koch, M\"uller and Rafelski
\cite{Rafelski}.

In order to define an overall chemical equilibration time
we perform a fit to
the particle abundances $N(t)$ for pions and kaons as
\begin{eqnarray}
N(t) = N_{eq} \left(1 - \exp(-t/\tau_{eq})\right),
\label{taueq}\end{eqnarray}
where $N_{eq}$ is the equilibrium limit.  The equilibration time $\tau_{eq}$
thus corresponds to the time $t$ when $\simeq 63$\% of $N_{eq}$ is achieved.

\begin{figure}[t]
\centerline{\psfig{figure=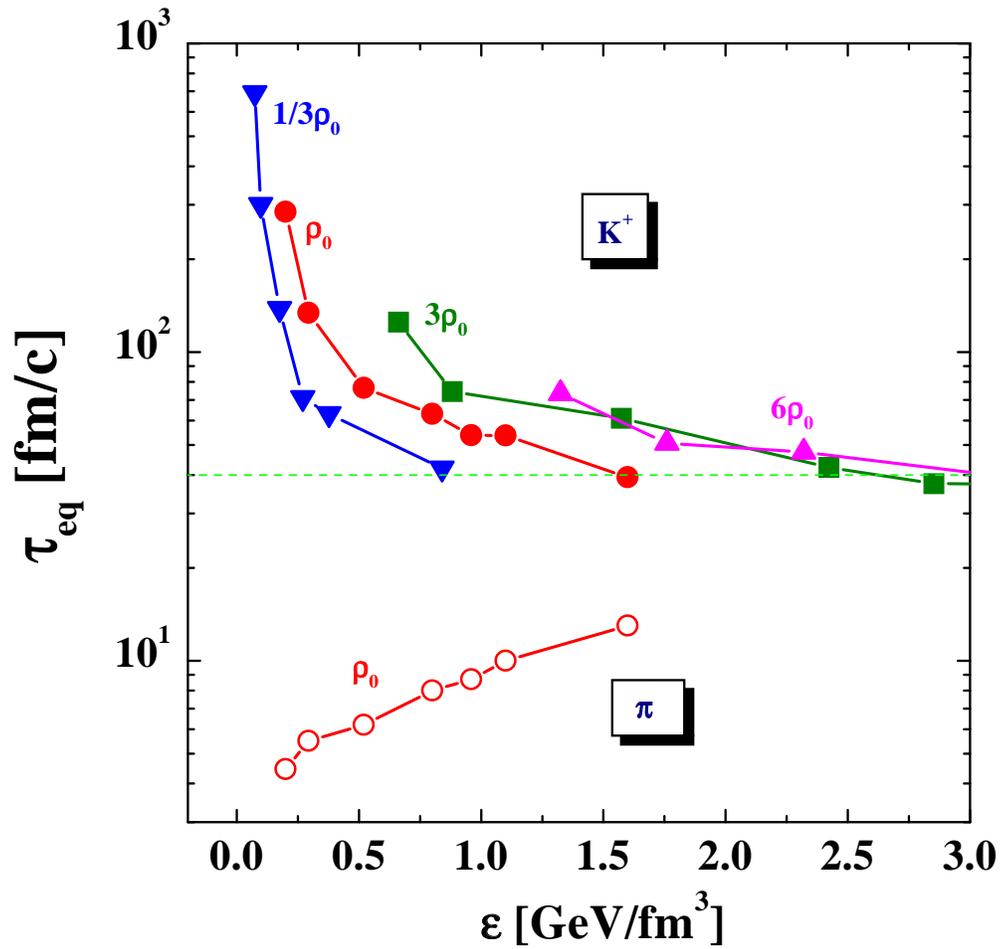,width=16cm}}
\vspace*{-2.5cm}
\caption{Equilibration time $\tau_{eq}$ versus energy density $\varepsilon$
for $\pi$ and $K^+$ mesons at different baryon densities $1/3\rho_0,
\rho_0, 3\rho_0$ and $6\rho_0$.}
\label{Fig4}
\end{figure}

Figure \ref{Fig4} shows the equilibration time $\tau_{eq}$ versus
energy density for $\pi$ and $K^+$ mesons at different baryon densities
of $1/3\rho_0, \rho_0, 3\rho_0$ and $6\rho_0$. We find that the
equilibration time for pions scales as $\tau_{eq}^\pi \sim 1/\rho$ or
$\Gamma_\pi \sim \rho$, thus we present the curve only for baryon
density $\rho_0$.  Whereas $\tau_{eq}^\pi$ slowly grows with
energy-density, $\tau_{eq}^K$ falls steeply with $\varepsilon$. This
marked difference is due to the fact that, on one hand, the kaon
production rate increases dramatically with $\sqrt{s}$ whereas that of
the pions, on the other hand, is more flat. With increasing energy thus
more strange particles are produced through strings especially from the
primary collisions with high $\sqrt{s}$ and the chemical equilibration
is achieved faster.

In Fig.~\ref{Fig4} we have considered an 'ideal' situation, i.e. hadron
matter at fixed energy and baryon density.  In realistic heavy-ion
collisions the system goes through the different stages due to
interactions and expansion. However, as follows from Fig. \ref{Fig4},
the equilibration time for strangeness is larger than 40~fm/c for all
energy and baryon densities. Thus in realistic nucleus-nucleus
collisions the chemical equilibration of strange particles requires
also a time above~40 fm/c which is considerably larger than the actual
reaction time of a few 10 fm/c or less (cf. Section 4).

The particle abundances used to extract $\tau_{eq}$ in Fig. \ref{Fig4}
have been calculated without any in-medium potentials.
In fact, the introduction of attractive potentials (especially for
$K^-$) will lower the hadronic thresholds and thus increase the
scattering rate between strange and nonstrange hadrons, whereas the
$K^+$ feels some repulsive potential and the trend goes in the opposite
way.  According to our calculations such in-medium modifications (in
line with Ref. \cite{Cass99}) give a correction to the $K^+$
equilibration times by atmost 10 \% and shortens the $K^-$
equilibration times up to 20 \% at density $\rho_0$.

\subsection{Thermal equilibration and limiting temperature}

In this subsection we investigate the approach to thermal equilibration.
This is initially driven by the very early string phase on the momentum
equilibration of the hadronic degrees of freedom, when the system is
still very far from equilibrium and the energy density is sufficiently
high. This one can see by looking at the quadrupole moment
${<}Q_2{>}=<2p_z^2-p_x^2-p_y^2>$ of the momentum distribution of all hadrons
involved. In the left panel of Fig.~\ref{Fig5} we present the time
evolution of the quadrupole moment $<Q_2>$ for density $\rho=\rho_0$ at
energy densities $\varepsilon=0.22$, 0.3, 0.52, 0.8, 1.1 and 1.6
GeV/fm$^3$. In order to take into account the string contributions we
have counted here all particles even within the formation time.  The
thin solid lines indicate exponential fits of the form
\begin{eqnarray}
<Q_2>(t) \simeq A_1\exp(-t/\tau_{short})+ A_2\exp(-t/\tau_{long})
\label{Q2}\end{eqnarray}
with two equilibration times $\tau_{short}$ and $\tau_{long}$.

The right panel of Fig.~\ref{Fig5} shows $\tau_{short}$ and
$\tau_{long}$ versus energy density $\varepsilon$. Whereas
$\tau_{short}\simeq 5$~fm/c is roughly independent on $\varepsilon$ the
'hadronic' equilibration time $\tau_{long}$ increases with energy
density. These results have to be interpreted as follows: in the
initial nonequilibrium phase the string degrees of freedom are excited
and decay according to many-body phase on a short time scale
$\tau_{short}$. The string decays reduce the initial quadrupole moment
(at high energy density) in time by a significant factor of about
$3-4$. One can understand the result obtained for $\tau_{short}$ in a
rather simple way.  Due to our prescription of the initialization of
the system the first strings on average are formed  after the time
$\tau_{coll} \approx 1/((\rho/2) \sigma_{NN } \langle v_{NN} \rangle)
\approx 3-4$~fm/c for $\rho = \rho_0$.  The strings then decay within
their formation time $\tau_F \approx 0.8$~fm/c  giving rise to a
significant production of transversal momentum.  One should point out,
that according to these arguments $\tau_{short} $ approximately scales
like $1/\rho $.  Due to Lorentz contraction $\tau_{short}$ is thus
considerably smaller in a real heavy-ion collision. Hence, string
decays provide a very efficient source for a strong decrease in
longitudinal momentum and production of transverse momentum in the very
early stage of an ultrarelativistic heavy-ion collision. A decrease
(increase) of the formation time $\tau_F$ to 0.5 fm/c (1.5 fm/c)
changes $\tau_{short}$ on the scale of 20\%, only.

After string decay, however, the emerging hadronic system still has
significantly larger longitudinal than transverse momenta -- the ratio
increases with energy density $\varepsilon$ -- and low energy hadronic
reactions are less effective in transfering longitudinal to transverse
momentum or simply in production of mesons.  This explains the increase
of $\tau_{long}$ with $\varepsilon$ in simple terms.

 From the above analysis it follows that after typical relaxation times
of $\tau_{short} \approx$ 5~fm/c the momentum unisotropy
of hot and dense matter has dropped to ${\rm e}^{-1}$ such that one might
describe the system by simple global thermodynamical variables like
temperature etc. This thermal equilibrium has to be contrasted with the
chemical equilibrium which -- as we have shown in the preceding
subsection -- is reached only after much longer times ($\geq 40$ fm/c
for strange particles, for example).

For the equilibrated system we can extract a temperature $T$ by fitting the
particle spectra with the Bolzmann distribution
\begin{eqnarray}
{d^3N_i\over dp^3} \sim \exp(-E_i/T),
\label{Boltz}\end{eqnarray}
where $E_i=\sqrt{p_i^2+m_i^2}$ is the energy of particle $i$.  We note
that at the temperatures of interest here, the Bose and Fermi
distributions are practically identical to a Boltzmann distribution. We
find that in equilibrium  the spectra of all particles can be
characterized by one single temperature $T$. This is demonstrated in
Fig.~\ref{Fig6} where we show the spectra of nucleons ($N$), pions ($\pi$)
and kaons ($K^+$) as a function of the kinetic energy $E-m$ for
$\rho=\rho_0$ at energy densities $\varepsilon=0.52, 0.8$ and 1.6
GeV/fm$^3$ (left panel) and for $\rho=3\rho_0$ at energy densities
$\varepsilon=0.66, 1.57$ and 2.85 GeV/fm$^3$ (right panel). Here we have
averaged the spectra from 950 fm/c to 1000 fm/c in order to decrease the
numerical fluctuations. The spectra of $N, \pi, K^+$ here can be
fitted with a single temperature $T$ which increases with the energy
density $\varepsilon$ for both baryon densities $\rho_0$ and $3\rho_0$.
We note explicitly that the slope of the equilibrium particle spectra
does not depend on the formation time $\tau_F$.

In Fig.~\ref{Fig7} we display the energy density $\varepsilon$ versus
temperature $T$ for different baryon densities $\rho$:  $1/3\rho_0$
(open down triangles), $\rho_0$ (full squares), $3\rho_0$ (full dots),
$6\rho_0$ (full up triangles). In order to compare calculations for
different baryon densities we have subtracted the baryon energy density
at rest, i.e. $\simeq m_N\rho$ (except for Fermi motion). As seen from
Fig.~\ref{Fig7} the temperature grows with energy density up to a
limiting value reminiscent of a 'Hagedorn' temperature \cite{Hagedorn}.
 From our detailed investigations we obtain for the limiting temperature
$T_s \simeq 150\pm 5$~MeV which practically does not depend on baryon
density.  Such a singular behavior of $\varepsilon(T)$ for $T\simeq
T_s$ has also been found in the box calculations in Ref.~\cite{Brav1}
for $\rho=\rho_0$.  Our limiting temperature is slightly higher than
that in Ref.~\cite{Brav1} ($T_s = 130 \pm 10$~MeV)  due to the different
number of degrees of freedom; the model \cite{Brav1} contains more
resonances and uses a different threshold for string excitations.
Thus, there is some phenomenological sensitivity to the hadronic zoo of
particles and string thresholds employed in the model.

In Fig. \ref{Fig8} we show the excitation function for the ratio of
string energy density to the energy density of the whole system
$\varepsilon_{string}/\varepsilon$ at $\rho=\rho_0$ when the system has
equilibrated for long times.  If the equilibrated system is very dense,
lower energy strings are still continuously being excited and thus --
because of their subsequent decay -- the strings constitute a
stationary portion of the total energy of the system.  The relative
ratio in the energy density increases with $\varepsilon$ up to a
saturation value of $\simeq 16$\% and then stays essentially constant. This
reflects that the system reaches a limiting temperature, since the
relative amount of string excitations compared to resonance excitations
does not change any more, whereas the number of strings as well as the
number of hadrons produced increases with $\varepsilon$. This fact one
might have guessed since the string production rate in equilibrium
depends only on the temperature $T$ characterizing the Bose/Fermi
distributions in the collision terms. In addition, this constant
fraction, of course, also intrinsically depends on the excitation
threshold and on the chosen decay (or formation) time $\tau_F$ of the
strings.

As pointed out above, the string degrees of freedom play an essential
role for particle production at high bombarding energies since they
describe the continuum excitations of the system.  The number of
strings created is especially high at the first stages of the
collision, when the energy of baryon-baryon interactions is close to
the initial energy $\sqrt{s}$.  It decreases with time to some constant
value which corresponds to the equilibrium state. Because of this
string-dominance one now has to worry about possible consequences of a
violation of detailed balance for these degrees of freedom. As already
pointed out earlier, all hadronic cascade-type approaches use the
phenomenological string picture in order to describe quantitatively
energetic (soft or semi-hard) inelastic reactions above some specified
$\sqrt{s}$-threshold. In such binary hadronic reactions typically many
hadronic particles and resonances are produced, the number depending on
the incident energy $\sqrt{s}$. The `back reaction' of these particles
produced from decay of an excited string (or two strings in the LUND
model) leading to the formation of only two energetic hadrons again is
not considered as it is statistically suppressed  and difficult to
describe. On the other hand, in an `infinite' matter calculation these
back reactions have to be taken into account in order to
allow for the principle of detailed balance. This is not done here as
it is technically difficult to handle; it thus represents a potential
`Achilles heal' in a thermodynamic analysis. However, for simulating a
heavy-ion collision this deficiency is not of any major importance
since the excitation of strings happens in the first moment of the
reaction when the phase space is still widely open and no back reaction
can occur.

\subsection{Comparison to the statistical model}

In order to investigate the equilibrium behavior of hadron matter we
also compare our transport (box) calculations with a simple Statistical
Model (SM) for an Ideal Hadron Gas (IHG) where the system is described
by a grand canonical ensemble of non-interacting fermions and bosons in
equilibrium at temperature $T$.  All baryon and meson species
considered in the transport model \cite{Effe99gam} also have been
included in the statistical model. Our main objective here is to
compare our results with the Hagedorn bootstrap picture of hadronic
matter \cite{Hagedorn}.

We recall that in  the SM particle multiplicities $n_i$ and energy
densities $\varepsilon_i$ are given by
\begin{eqnarray}
&& n_i ={g_i \over (2\pi \hbar)^3} \int\limits_0^\infty
{4\pi p^2 dp \over \exp\left[(E_i - B_i\mu_B - S_i \mu_S)/T\right]\pm 1},
 \label{Nth} \\
&& \varepsilon_i ={g_i \over (2\pi \hbar)^3}
\int\limits_0^\infty {4\pi E_i p^2 dp \over
\exp\left[(E_i - B_i\mu_B - S_i \mu_S)/T\right]\pm 1},
\label{Enth}
\end{eqnarray}
where $E_i = \sqrt{p^2+m_i^2}$ is the energy of particle $i$, $B_i$ is
the baryon charge, $S_i$ is the strangeness, and $g_i$ is the
spin-isospin degeneracy factor.  In Eqs. (\ref{Nth}),(\ref{Enth})
$\mu_B$ and $\mu_S$ are the baryon and strangeness chemical potentials.
Here we neglect the electric chemical potential ($\mu_n=\mu_p=\mu_B$)
since we consider an isospin symmetric system.  Note, however, that in
realistic collisions of heavy-ions (like Au~+~Au) this reduction is no
longer fully appropriate.
For particles with finite spectral width we include in Eqs.
(\ref{Nth}),(\ref{Enth}) the spectral functions $\rho_i(m) $ with the
same parametrization for the width as in the transport model,
\begin{eqnarray}
&& n_i ={g_i \over (2\pi \hbar)^3} \int \rho_i(m) dm
\int\limits_0^\infty
{4\pi p^2 dp \over \exp\left[(E_i - B_i\mu_B - S_i \mu_S)/T\right]\pm 1},
\label{Nthsf} \\
&& \varepsilon_i ={g_i \over (2\pi \hbar)^3} \int \rho_i(m) dm
\int\limits_0^\infty {4\pi E_i p^2 dp \over
\exp\left[(E_i - B_i\mu_B - S_i \mu_S)/T\right]\pm 1}.
\label{Enthsf}\end{eqnarray}

The energy density $\varepsilon$, baryon density $\rho$ and strange
density of the hole system in equilibrium then given as
\begin{eqnarray}
&& \varepsilon =  \sum\limits_i \varepsilon_i (T,\mu_B,\mu_S)  \label{3eq_en} \\
&& \rho = \sum\limits_i B_i \ n_i (T,\mu_B,\mu_S) \label{3eq_rhoB} \\
&& \rho_S = \sum\limits_i S_i \ n_i (T,\mu_B,\mu_S)
\, \equiv \, 0 . \label{3eq_rhoS}
\end{eqnarray}

As 'input' for the SM we use the same $\varepsilon, \rho$ and $\rho_S$
as in the box calculations and  we obtain the thermodynamical
parameters -- $T, \mu_B, \mu_S$ -- by solving the system of nonlinear
equations (\ref{3eq_en}),(\ref{3eq_rhoB}) and (\ref{3eq_rhoS}).

Within the SM we find that the temperature increases continuously
with energy density since the continuum excitations, i.e. the string degrees
of freedom, are not included (full dots  in Fig.~\ref{Fig9}),
whereas the box calculation with strings gives the limiting temperature
(full squares in Fig.~\ref{Fig9}).  Both curves in Fig.~\ref{Fig9}
have been calculated for density $\rho_0$.

To reproduce qualitatively our box result within the SM we have to include
continuum excitations in the statistical model, i.e.
a Hagedorn mass spectrum for strings as defined by \cite{Hagedorn}
\begin{eqnarray}
\rho^{str}(m) = {\rho_0^{str} \over m^3} \exp(m/T_H),
\label{Hagden}\end{eqnarray}
where $T_H$ denotes the 'Hagedorn' temperature. For $T_H$ we use the
temperature $T_s$ as obtained from the box calculations, i.e.
$T_H=T_s\simeq 150$~MeV. In (\ref{Hagden}) $\rho_0^{str}$ is a fit
parameter additionally to  $T, \mu_B$ and $\mu_S$ to reproduce
$\varepsilon(T)$ from the box calculations. The string multiplicities
$n_i^{str}$  are given by
\begin{eqnarray}
&& n_i^{str} ={1 \over (2\pi \hbar)^3} \int\limits_{m_{min}}^\infty
\rho_i^{str}(m) dm \int\limits_0^\infty
{4\pi p^2 dp \over \exp\left[(E_i - B_i\mu_B - S_i \mu_S)/T\right]\pm 1},
\label{Nstr} \end{eqnarray}
where the lowest mass in the string excitation ($m_{min}$) is defined
by the string threshold in the transport model:  $m_{min}=2.6-m_N$~GeV
for baryon strings and $m_{min} =2.2$~GeV for meson strings.  In our
transport model we include the following strings $i$:  baryon strings
$B=1,S=0,-1,-2,-3$, anti-baryon strings $B=-1, S=0,1,2,3$, meson
strings $B=0, S=0,1$ and anti-meson strings $B=0, S=-1$.

Before going over to the actual analysis we point out that the limiting
temperature $T_s$ from our string model involves somewhat different
physics assumptions than the Hagedorn model at temperature $T_H$. $T_s$
should not really be identified with the 'Hagedorn' temperature $T_H$,
though close similarities exist. In the Hagedorn picture and for
temperatures close to $T_H$ the abundance of `normal' hadrons or known
resonances stays constant with increasing energy density whereas the
number and energy density of the (hypothetical) bootstrap excitations
diverges for $T\rightarrow T_H$. The Hagedorn model thus assumes
`particles' of mass $m\to\infty$ to be populated for $T\to T_H$, that
dynamically can be formed in collisions of high mass hadrons for
$t\to\infty$.  In contrast, our string model does not include energetic
string-string interactions that might produce more massive strings.
(There exist some phenomenological recipes how to incorporate such
interactions \cite{Sailer}.) The 'high mass' strings decay to hadrons
and, because of the detailed balance problem discussed in the last
subsection, are only repopulated by binary hadron-hadron or
hadron-string interactions, so that their internal energy is limited
and the low-energy hadronic degrees of freedom are overpopulated. This
leads to the saturation of string-energy to total energy (observed in
Fig.~\ref{Fig8}) to a value of $\simeq 0.16$ in contrast to the value
of 1 in the Hagedorn model.

This, however, does not imply a fundamental inconsistency for the
overall properties of the system. In perfect chemical equilibrium, like
in the Hagedorn model, more strings (or hypothetical resonances) would
be excited which, for lower temperatures (e.g.\ in a nearly isentropic
expansion of the system like in heavy-ion collisions), would
immediately decay into a large number of hadronic particles. The
violation of detailed balance in our case thus physically describes an
overpopulation of hadronic particles only in stationary equilibrium.
The important point, however, is the observation that in either
description the system at equilibrium can not exceed the critical
temperature $T_s$.

As seen in Fig.~\ref{Fig9} we achieve agreement of the extended SM and
our box calculations from Fig.~\ref{Fig8} by choosing $T_H \approx T_s$
in Eq.\ (\ref{Hagden}). In addition, from the extended SM we can also
define thermodynamical parameters such as the baryon chemical potential
$\mu_B$. In Fig.~\ref{Fig10} we present the resulting $T-\mu_B$
correlation, i.e. temperature $T$ versus baryon chemical potential
$\mu_B$, at fixed baryon densities (in the box calculations) of $\rho =
1/3\rho_0, \rho_0$ and various energy densities. The open triangles and
squares (connected by the dashed lines) show the result of the SM
without strings at densities $1/3\rho_0$ and $\rho_0$, respectively,
whereas the full triangles and squares (connected by the solid lines)
correspond to the thermodynamical fit of the box calculations (at
$1/3\rho_0$ and $\rho_0$) including string excitations.
The errorbars indicate the uncertainty in the extraction of $\mu_B$ in
the SM; they become larger when the system is closer to $T_H$ due to
the divergence in the energy density integral (\ref{3eq_en}). The
arrow at $\mu_B=0$  indicates the temperature $T_s=150$~MeV from
our box calculations. The full dots in Fig.~\ref{Fig10} correspond to
chemical freeze-out points extracted in a thermodynamical model from
hadron abundances \cite{BM}; the open dots are the thermal
freeze-out points from the momentum spectra of hadrons and two-particle
correlations as taken from Ref.~\cite{HeinzQM97}.

Our calculations here are for nuclear matter densities $1/3\rho_0$ and
$\rho_0$ whereas the freeze-out points have been extracted from heavy-ion
data; the comparison thus can be only qualitative.  However, one can see
the general tendency: if the continuum excitations (strings) are not
included in the thermodynamical analysis, one can 'extract' much larger
temperatures at high energy density simply due to the limited number of
degrees of freedom involved in the model analysis. In this respect our
box result  is more in line with the thermal ('kinetic') freeze-out
analysis from Ref.~\cite{HeinzQM97} than with the thermodynamical
analysis from Ref.~\cite{BM} that is based on particle ratios and thus
on chemical freeze-out. The point to make is that
at higher temperatures, like e.g. the ones obtained
for a `chemical' freeze-out in Ref.~\cite{BM},
the consideration of continuum excitations does make a thermodynamical
analysis much less certain than at lower temperatures, like e.g.
at `thermal' (or kinetic) freeze-out as in Ref. ~\cite{HeinzQM97},
where the continuum excitations do not play any significant role.

In this context we have to mention, furthermore, that a combined
experimental analysis of particle spectra and HBT radii favors
even lower freeze-out temperatures (below 100 MeV \cite{Nix,Heinz99}).
For these freeze-out conditions the pion density (for fixed charge)
drops below $\sim 10^{-2}$~fm$^3$, i.e. the average distance between
two pions (of different charge) becomes large than $\sim 4.6$~fm,
which in turn is large compared to their classical interaction radius
$r_I=\sqrt{\sigma_{\pi\pi}/\pi}$ at all relative momenta between the
two pions.  Since thermal freeze-out temperatures of 90-100 MeV at SPS
energies can be considered as a lower bound, the 'experimental' points
in Fig.~\ref{Fig10} have to be taken with care.

\section{Expanding hadronic fireballs}

In realistic nucleus-nucleus collisions the system rapidly expands
after the possible formation of a hot hadronic fireball.  The final
hadronic spectra can be changed substantially during this expansion
phase, i.e.  the temperature extracted from the experimentally observed
slopes of the spectra also contains information about the nuclear
expansion dynamics.

To investigate the expansion of the hadronic fireball we initialize the
system in a box with periodic boundary conditions -- as described above
-- and propagate the system up to 500 fm/c, when equilibrium is
reached. Afterwards we let the system expand without boundary
conditions. Even though this is an idealized description of the
expansion phase during a heavy-ion collision we hope to learn from this
scenario how the expansion stage changes the picture of perfect thermal
equilibrium (for an analysis of an actual collision see the discussion
in the following section).

In Fig.~\ref{Fig11} we present the time evolution of the various
particle abundances (nucleons $N$,  $\Delta$,  $\Lambda$, $\pi$, $K^+$
and $K^-$ mesons) during the expansion for density $\rho=\rho_0$ (left
panel) at different energy densities $\varepsilon =0.22, 0.3$ and 1.1
GeV/fm$^3$ and for density $\rho=1/3\rho_0$ at $\varepsilon =0.84$
GeV/fm$^3$ (upper part in the right panel), for $\rho=\rho_0$ at
$\varepsilon =1.6$ GeV/fm$^3$ (middle part in the right panel) and for
$\rho=3 \rho_0$ at $\varepsilon =3.4$ GeV/fm$^3$ (lower part in the
right panel).
The number of stable particles ($N, \Lambda, \pi, K^+, K^-$) increases
during the expansion up to some asymptotic value due to string and
heavy resonance decay as well as inelastic interactions.
One can see that the asymptotic values are reached after a few 10 fm/c from
the beginning of the expansion (depending on the initial
energies and baryon densities) which is comparable to the actual
reaction time in heavy-ion collisions (cf. Section 4).

In Fig.~\ref{Fig12}  we show the spectra of nucleons ($N$), pions
($\pi$) and kaons ($K^+$) versus the kinetic energy $E-m$ for
$\rho=\rho_0$ at energy densities $\varepsilon=0.22, 0.3$ and
1.1~GeV/fm$^3$  before the expansion -- averaged over time from 450
fm/c to 500~fm/c -- (left panel) and after the expansion -- averaged from
580 fm/c to 600~fm/c -- (right panel). For completeness in
Fig.~\ref{Fig13} we present the result for $\rho=1/3\rho_0$ at
$\varepsilon=0.84$ GeV/fm$^3$ (upper part), for $\rho=\rho_0$ at
$\varepsilon=1.6$ GeV/fm$^3$ (middle part) and for $\rho=3\rho_0$ at
$\varepsilon=3.4$ GeV/fm$^3$ (lower part). In the left panels the
systems are in equilibrium; the $N, \pi, K^+$ spectra show a common
temperature $T$ whereas after the expansion the slopes of the particle
spectra are different; the nucleon  spectra are much harder than the
pion spectra, i.e. the apparent temperature of particles (after the
expansion) increases with the mass $m$.  This effect is illustrated in
Fig.~\ref{Fig14}, where we show the apparent slope $T$ versus $m$ for
$\pi, K^+, N$ for $\rho=\rho_0$ at different energy densities:
$\varepsilon = 0.2$~GeV/fm$^3$ (full up triangles), $\varepsilon =
0.3$~GeV/fm$^3$ (full squares), $\varepsilon = 1.1$~GeV/fm$^3$ (full
dots), $\varepsilon = 1.6$~GeV/fm$^3$ (full diamonds); and for
$\rho=3\rho_0$ at $\varepsilon=3.4$~GeV/fm$^3$ (open down triangles).
The arrow indicates the limiting temperature $T_s=150$~MeV before the
expansion.  One can see from Figs.~\ref{Fig12}--\ref{Fig14} that the
'expansion' temperature of particles increases also with the energy
density.  However, if during the equilibration phase the system reaches
$T_s$, the 'expansion' temperatures for different particles show a
universal behaviour, i.e. practically do not depend on the energy
$\varepsilon$ as well as on the baryon density $\rho$. This phenomenon
is due to the fact that close to $T_s$ the initial hadron velocity
distributions, reflected in the particle momentum profile, become
similar for all $\varepsilon$ and $\rho$ in equilibrium.

In order to investigate the origin of the enhancement in the particle slope
during the expansion we have performed several illustrative calculations:
at 500 fm/c -- after the system has achieved equilibrium -- we
i) let all resonances and strings decay; in this case we find that the
slopes do not change as compared to the equilibrium phase,
ii) we let the system expand without interactions (allowing only decays)
and find that the slopes slightly decrease in comparison to the
equilibrium phase.
Both examples indicate that the slope enhancement stems basically from
multiple interactions of the particles in the initial stages of
the expansion phase.

For analyzing the expansion flow phenomena we have performed a fit of
the particle spectra (after expansion) using the blast model of Siemens
and Rasmussen \cite{Siemens}. In this model all particle spectra are
described by a universal formula with common thermal freeze-out
parameters, i.e. a temperature $T$ of the fireball and a radial-flow
velocity $\beta$:
\begin{eqnarray}
{d^3N_i\over dp^3} = A_i \exp\left(-{\gamma E_i\over T}\right)
\left[{\sinh\alpha_i\over \alpha_i} \left(\gamma+{T\over E_i}\right) -
{T\over E_i}\cosh\alpha_i\right], \label{flow}
\end{eqnarray}
where $\gamma=(1-\beta^2)^{-1/2}, \ \alpha=\gamma \beta p_i/T$. Here
$E_i, p_i$ are the total energy and momentum of the considered particle
$i$ while $A_i$ are normalization factors.

We now try to describe the final particle spectra after the expansion
by Eq.~(\ref{flow}) with common freeze-out parameters $T$ and $\beta$.
In Fig.~\ref{Fig15} we show the result of our least-squares fit, using
the MINUIT method \cite{MINUIT}, for the energy densities $\varepsilon
=0.3$~GeV/fm$^3$ (upper part) and 1.1~GeV/fm$^3$ (lower part) at
$\rho=\rho_0$.  The left panel shows the contour plots for the
parameter errors in the $T-\beta$ plane (for the $\chi_{optimal}^2+1$
level); the dot-dashed lines stand for
nucleons ($N$), the solid lines for pions ($\pi$) and the dashed lines
for kaons ($K^+$). The full symbols indicate the 'best' values for $T$
and $\beta$ according to the $\chi^2$ criteria (squares for $N$, dots
for $\pi$ and triangles for $K^+$). The thin solid lines in the right
panel demonstrate the fit of the particle spectra within the optimal
parameters from MINUIT.

Since the particle spectra cover  several orders of magnitude and the
low energy points contribute to $\chi^2$ with a larger weight than
those at high energy, we use the logarithmic $\chi^2$ method to give a
higher weight to the tail of the spectra in the fitting procedure,
i.e.\  we minimize $\chi_{\ln}^2=\sum\limits_i (\ln f(x_i) - \ln
f_0(x_i))^2$, where $f_0$ represent the 'experimental' data (i.e. the
results of our box calculations), $f$ is the value of the fit
(\ref{flow}) at point $x_i$.

One can see from Fig.~\ref{Fig15} that the 'best' parameters $T$ and
$\beta$ (as well as the contours for the parameter errors) are
quite different for $N$, $\pi$ and $K^+$ especially for
$\varepsilon=0.3$~GeV/fm$^3$.  So we do not find (within the 'optimal'
$\chi^2$) common freeze-out parameters for all spectra simultaneously.
This is similar to an analysis of experimental spectra by Peitzmann et
al. \cite{Peitzmann}.  For all particles we obtain different
values for  $\beta$ and much lower temperatures $T$ than that of the
initial fireball:  $T_{in}=107$~MeV for $\varepsilon=0.3$~GeV/fm$^3$ and
$T_{in}=145$~MeV for $\varepsilon=1.1$~GeV/fm$^3$.

Since especially the pion spectra contain large contributions from
resonance decays at low $E-m$, we have also performed fits excluding
the pion spectra for $E-m \le 0.4$ GeV. This procedure essentially
gives lower $\beta$ parameters and higher values for $T$ (open circle
in the upper panel). However, the low energy cut-off is an additional
parameter that allows to 'extend' the ($\beta, T)$ values to a wider
range.

Thus our analysis indicates that the final particle spectra do not
allow a reliable reconstruction of freeze-out parameters within the
collective flow model (\ref{flow}). The parameters $T$ and $\beta$
obtained from the fit are very sensitive to the low energy shape of the
hadron spectra or low energy cut-off applied since this region
contributes with the largest weight to the $\chi^2$ minimization. On
the other hand, a global 'eye' fit with the parameters given by the
'star' for all hadrons considered gives a quite reasonable overall
description of the spectra (dashed lines, r.h.s of Fig.~\ref{Fig15}),
too. A very accurate deduction of one single overall fit for all
hadrons by a common temperature ($T$) and flow velocity ($\beta $)
parameter (see, e.g., \cite{HeinzQM97} and references therein) seems to
us thus rather ambiguous. In particular, such an analysis may indicate
that thermal equilibrium has been reached to a much larger extent than
is actually true.

The particle flow effect due to the expansion is demonstrated in
Fig.~\ref{Fig16} where we show the velocity distributions $dN/d\beta$
for nucleons ($N$), pions ($\pi$) and kaons ($K^+$) for $\rho=1/3 \rho_0$
at  $\varepsilon=0.84$~GeV/fm$^3$ (upper part), for $\rho=\rho_0$ at
$\varepsilon=1.1$~GeV/fm$^3$ (middle part) and for $\rho=3\rho_0$ at
$\varepsilon=3.4$~GeV/fm$^3$ (lower part).  The left panel shows the
$dN/d\beta$ distribution at equilibrium whereas the right panel
corresponds to $dN/d\beta$ after the expansion phase.  On can see that
the average velocity of the particles decreases with the mass; the
pions are much faster than the nucleons. They thus leave the reaction
zone  at the initial stage of the ongoing and rapidly evolving
expansion with a higher velocity and accelerate the slower hadrons that
'feel' the 'pion wind' by the multiple interactions \cite{Pang}.
We recall that the pion density is very high especially at high
$\varepsilon$ such that practically all other hadrons are shifted in
direction of large $\beta$. The same effect is shown in
Figs.~\ref{Fig12},\ref{Fig13} by the enhancement of the slope of the
nucleon spectra due to the expansion.

\section{Reactions of finite systems}

In this Section we turn to realistic nucleus-nucleus collisions with
the BUU transport model. We have learned from our analysis in the
previous Section that even by starting from an idealized scenario of
perfect thermal equilibrium, a rapid expansion stage makes the
extraction of one global temperature $T$ and one global expansion
parameter $\beta$ quite ambiguous. We thus expect this to become even
worse for the true situation of a relativistic heavy-ion collision,
where a perfect equilibrium state at some intermediate stage cannot
really be assumed. Also, we note that even for a very heavy system like
Pb+Pb the fraction of effective surface layer ($\sim 4\pi R_{eff}^2
\lambda$) to total volume is still quite sizeable, resulting in a
continuous emission or evaporation of particles from the outer layers
before a global freeze-out of bulk particles occurs. (For Pb+Pb
reactions at SPS energies -- combining a hydrodynamical evolution with
a nonequilibrium picture of surface emission -- it has indeed been
shown that at least 25$\%$ of all particles are continuously evaporated
before a global freeze-out has occurred \cite{Dumitru}.)

In Fig.~\ref{Fig17} we show the time evolution of the particle
abundances (nucleons $N$,  $\Delta$, $\Lambda$, $\pi$, $\eta$,  $K^+$)
for central collision of the light system $^{12}$C~+~$^{12}$C (upper
part) and heavy system $^{197}$Au~+~$^{197}$Au and
$^{208}$Pb~+~$^{208}$Pb (lower part) at the low energy of 1 A$\cdot$GeV
(left panel) and the high energy of 100 and 160 A$\cdot$GeV (right
panel).  The number of nucleons decreases in a few fm/c due to the
inelastic collisions, whereas the number of $\Delta$-resonances
increases accordingly. At low energy the $\eta$-mesons and strange
particles (we disregard strange particles for C~+~C at 1 A$\cdot$GeV
due to the low statistics) appear with a delay of a few fm/c due to the
fact that they are basically produced from resonance decays (the same
as pions) or from secondary pion-baryon collisions, whereas at high
energy they appear earlier due to the primary production mechanism
through the string formation and decay.

As seen from Fig~\ref{Fig17} the reaction time $\tau_{reac}$ for
1~A$\cdot$GeV is $\sim 20-30$~fm/c, whereas for high energies
$\tau_{reac}$ is shorter due to a faster expansion -- $\tau_{reac}
\simeq 10-20$~fm/c.  It has been shown in Section 2.2 that the chemical
equilibration of hadronic matter under 'ideal' conditions (box without
expansion) requires a quite long time, e.g. the equilibration time
$\tau_{eq}$ for strange particles has been found to be larger than 40
fm/c for all energies and densities (cf.  Fig.~\ref{Fig4}).  In
realistic central nucleus-nucleus collisions, such as Au~+~Au, the
system expands rapidly (depending on the energy) after the compression
and formation of the hadronic fireball. The number of interactions,
which is the dynamical origin for equilibration, decreases
correspondingly very fast with time; after a few 10 fm/c the
particles are moving practically freely. Thus, the reaction time even
for central Au~+~Au collisions is much shorter than the time required
for strangeness equilibration: $\tau_{reac} \ll \tau_{eq}$.

\begin{figure}[t]
\centerline{\psfig{figure=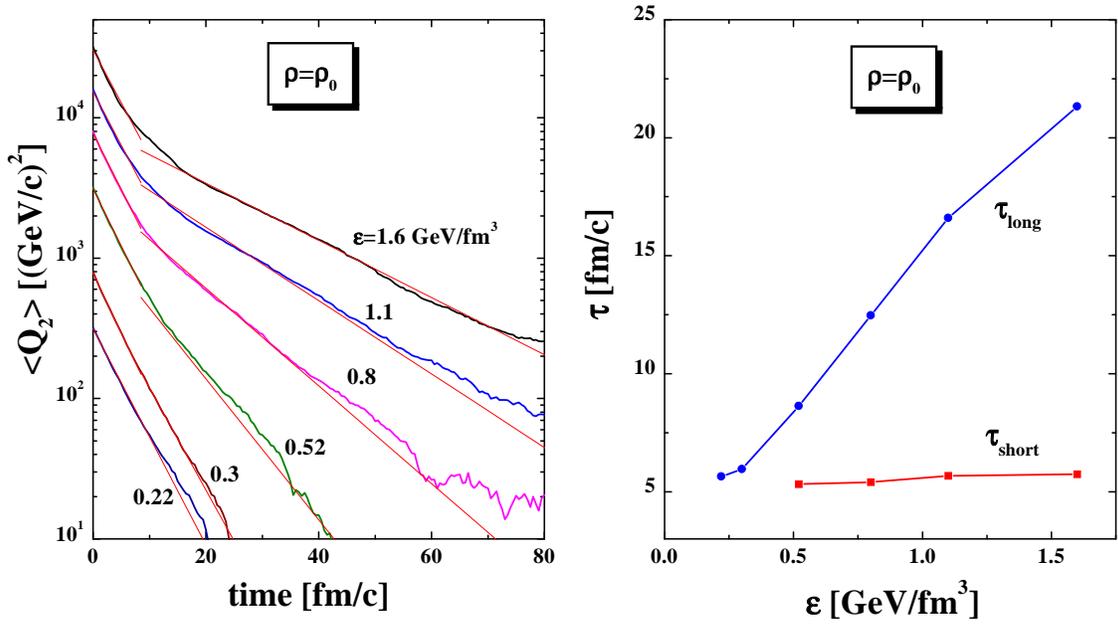,width=16cm}} \vspace*{-2.5cm}
\caption{Left panel: time evolution of the quadrupole moment $<Q_2>$
for density $\rho=\rho_0$ at energy densities $\varepsilon=0.22, 0.3,
0.52, 0.8, 1.1$ and 1.6 GeV/fm$^3$. The thin solid lines indicate the
fit of $<Q_2>$ by two exponentials (\ref{Q2}). The parameters
$\tau_{short}$ and $\tau_{long}$ are shown in the right panel as a
function of the energy density $\varepsilon$.}
\label{Fig5}
\end{figure}

\begin{figure}[t]
\centerline{\psfig{figure=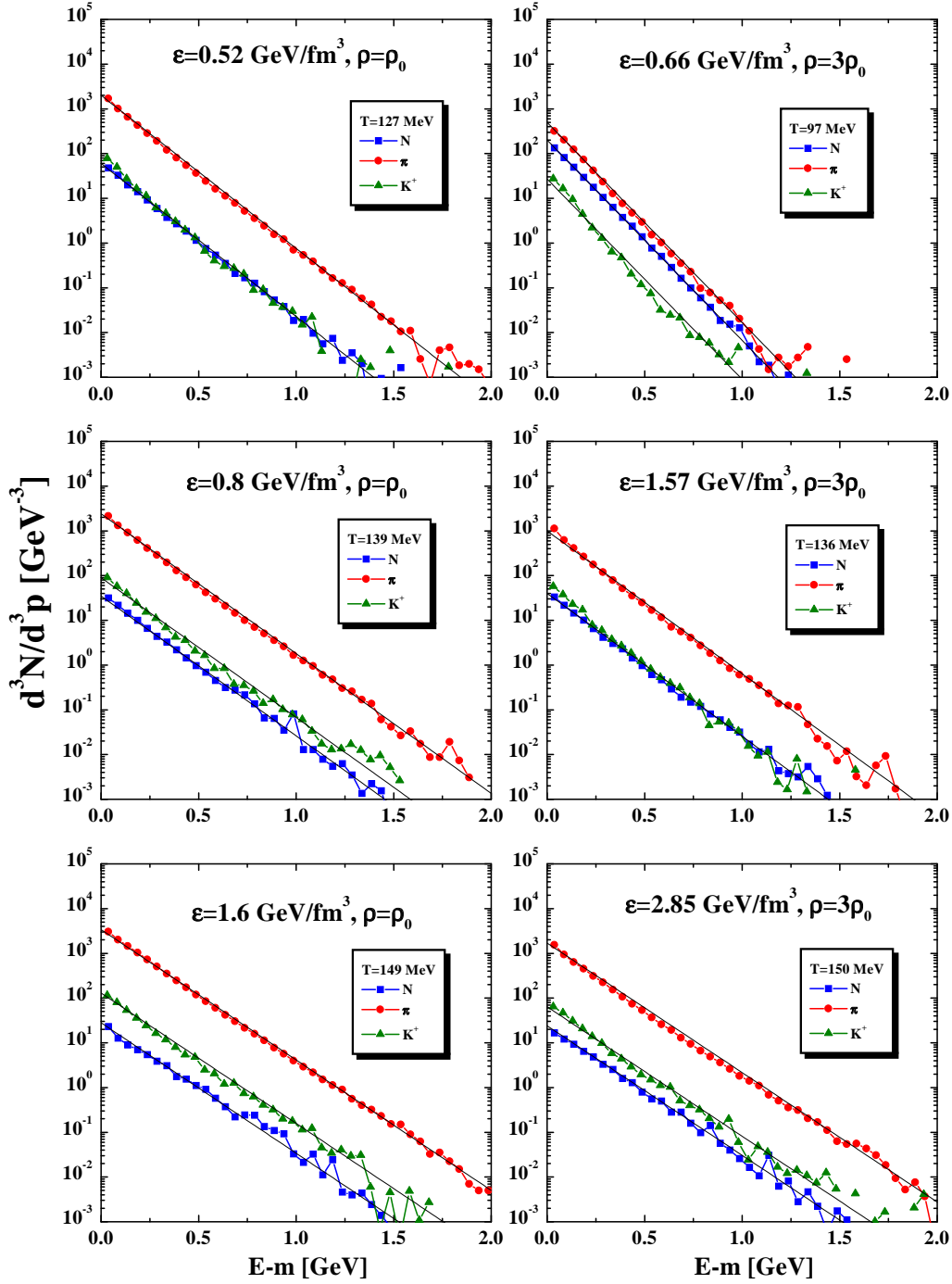,width=16cm}}
\vspace*{-2.5cm}
\caption{The spectra of nucleons ($N$), pions ($\pi$)
and kaons ($K^+$) as a function of the kinetic energy $E-m$ for
$\rho=\rho_0$ at energy densities $\varepsilon=0.52, 0.8$ and 1.6
GeV/fm$^3$ (left panel) and for $\rho=3\rho_0$ at energy densities
$\varepsilon=0.66, 1.57$ and 2.85 GeV/fm$^3$ (right panel).}
\label{Fig6}
\end{figure}

\begin{figure}[t]
\phantom{a}\vspace*{-2cm}
\centerline{\psfig{figure=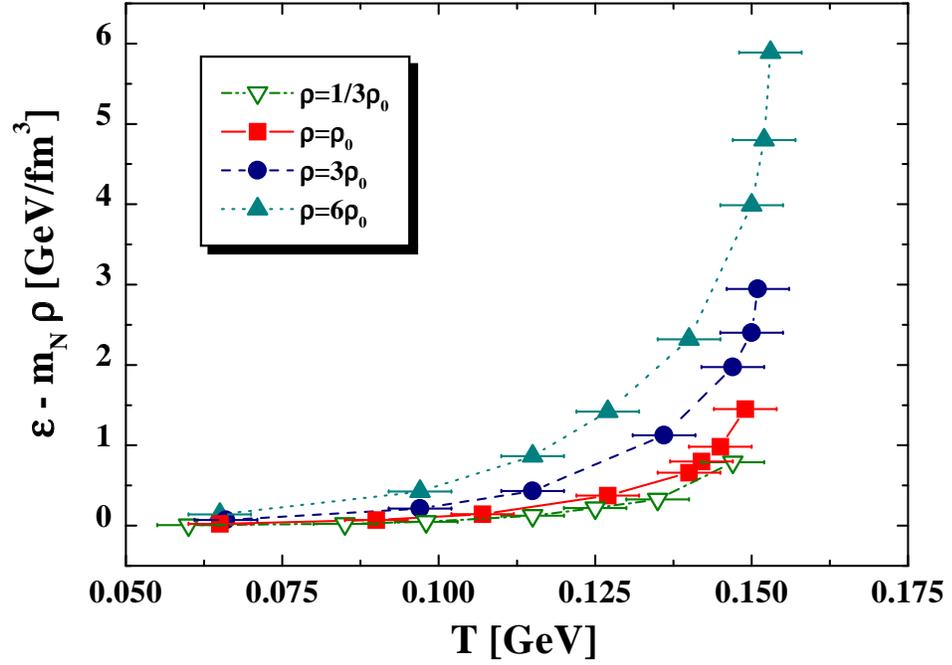,width=16cm}}
\vspace*{-1.5cm}
\caption{The energy density  $\varepsilon-m_N\rho$ versus equilibrium
temperature $T$ for different baryon densities $\rho$:  $1/3\rho_0$
(open down triangles), $\rho_0$ (full squares), $3\rho_0$ (full dots),
$6\rho_0$ (full up triangles). }
\label{Fig7}
\end{figure}

\begin{figure}[t]
\centerline{\psfig{figure=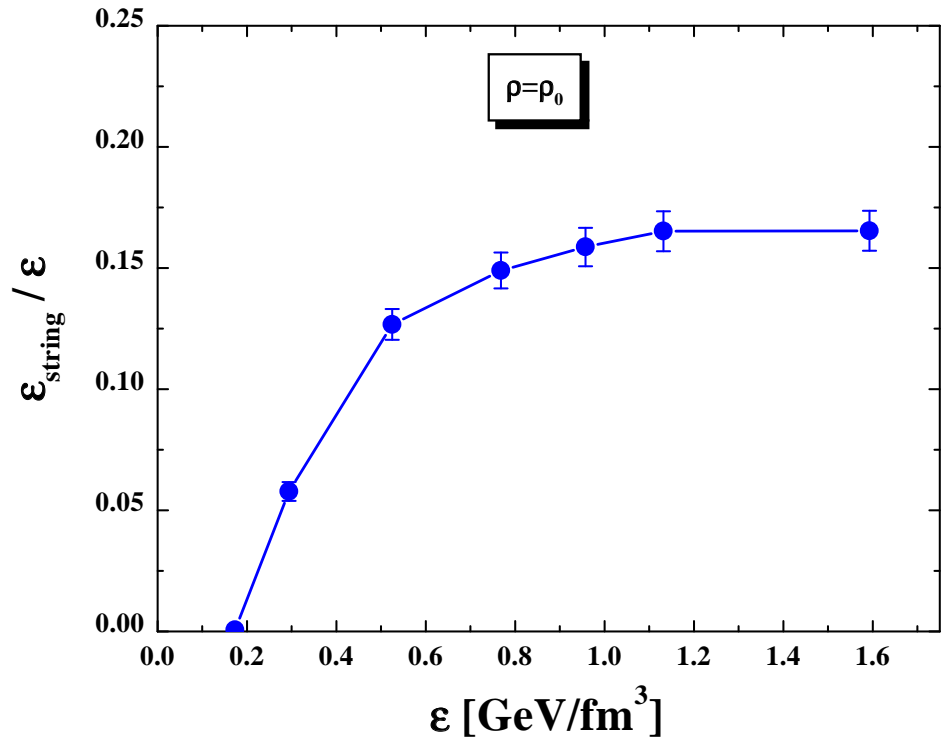,width=16cm}} \vspace*{-2.5cm}
\caption{The excitation function for the ratio of string energy density
to the energy density of the hole system
$\varepsilon_{string}/\varepsilon$ at $\rho=\rho_0$.}
\label{Fig8}
\end{figure}

\begin{figure}[t]
\phantom{a}\vspace*{-2cm}
\centerline{\psfig{figure=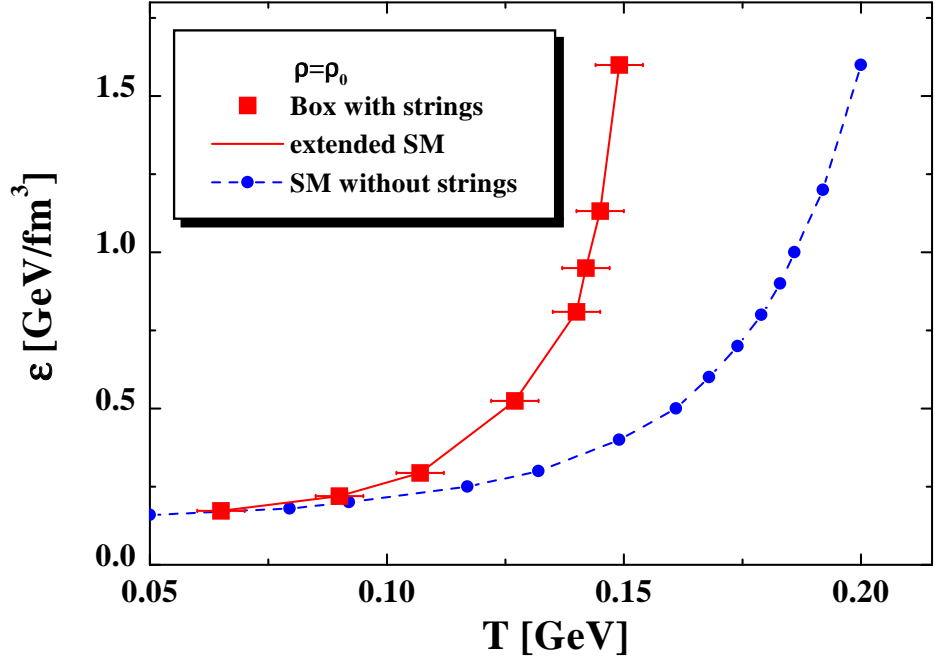,width=16cm}}
\vspace*{-1.5cm}
\caption{The energy density versus equilibrium temperature $T$ for
baryon density $\rho=\rho_0$. The full dots  correspond to the
statistical model (SM) without strings, the full squares show our
box calculations including string degrees of freedom, while the
solid line shows the result from the extended SM including a Hagedorn
mass spectrum for strings.}
\label{Fig9}
\end{figure}

\begin{figure}[t]
\phantom{a}\vspace*{-1.5cm}
\centerline{\psfig{figure=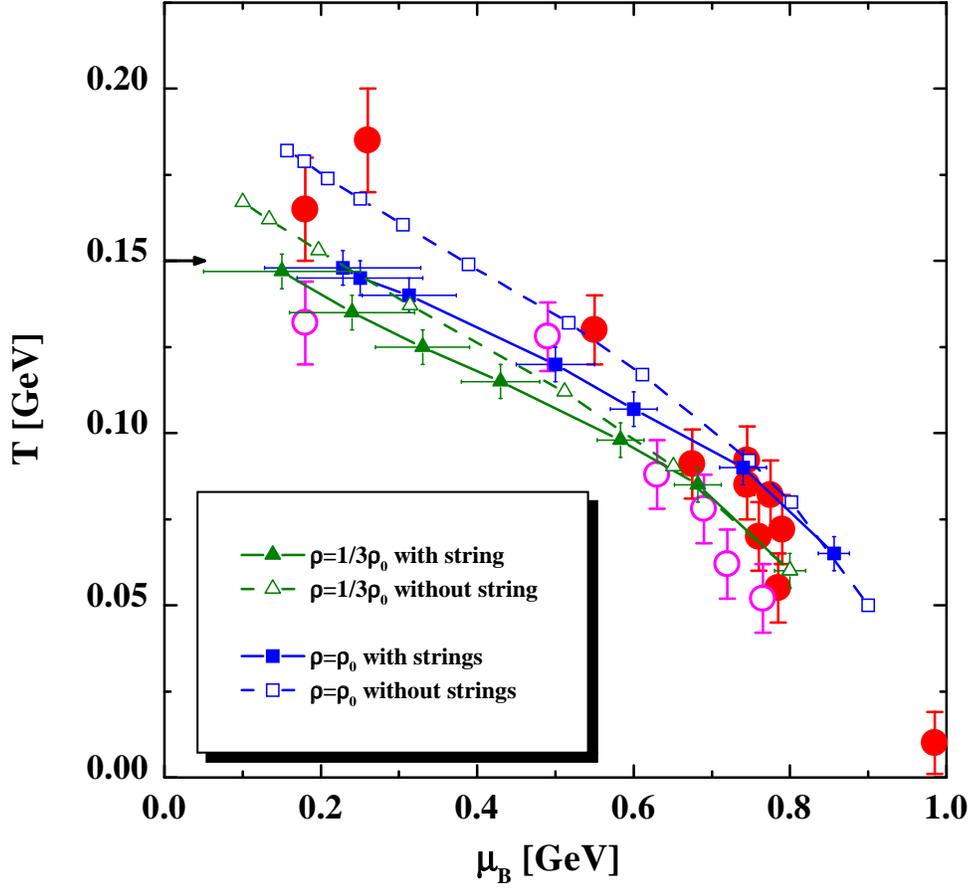,width=16cm}} \vspace*{-3.5cm}
\caption{The $T-\mu_B$ phase correlation, i.e. temperature $T$ versus
baryon chemical potential $\mu_B$. The open triangles and squares
(connected by the dashed lines) show the result of the statistical
model without strings (standard SM) fitted to out box calculations at
densities $1/3\rho_0$ and $\rho_0$, respectively, whereas the full
triangles and squares (connected by the solid lines) correspond to the
thermodynamical fit of the box calculations (at $1/3\rho_0$ and
$\rho_0$) including string excitations (extended SM). The arrow at
$\mu_B=0$ indicates the limiting temperature $T_s=150$~MeV from our box
calculations.  The full dots correspond to the chemical freeze-out
points from Ref.~\protect\cite{BM} while the open dots are the thermal
freeze-out points from Ref.~\protect\cite{HeinzQM97}. }
\label{Fig10}
\end{figure}

\begin{figure}[t]
\phantom{a}\vspace*{-0.5cm}
\centerline{\psfig{figure=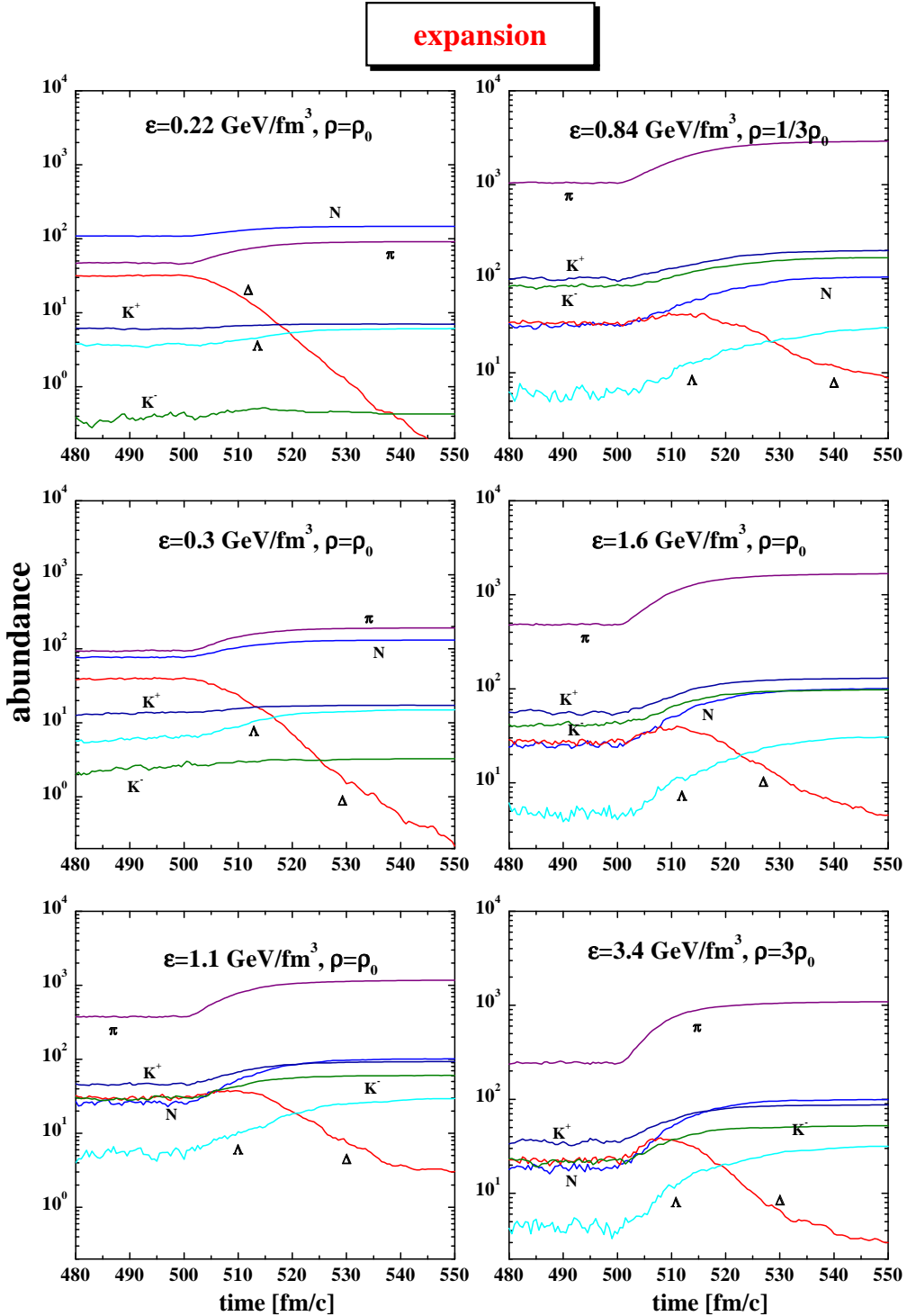,width=16cm}}
\vspace*{-2.5cm}
\caption{Time evolution of the various particle abundances (nucleons
$N$,  $\Delta$,  $\Lambda$, $\pi$, $K^+$ and $K^-$ mesons) during the
expansion (starting at $t=500$~fm/c) for density $\rho=\rho_0$
(left panel) at different energy densities $\varepsilon =0.22, 0.3$ and
1.1 GeV/fm$^3$ and for density $\rho=1/3\rho_0$ at $\varepsilon =0.84$
GeV/fm$^3$ (upper part in the right panel), for $\rho=\rho_0$ at
$\varepsilon =1.6$ GeV/fm$^3$ (middle part in the right panel) and for
$\rho=3 \rho_0$ at $\varepsilon =3.4$ GeV/fm$^3$ (lower part in the
right panel).}
\label{Fig11}
\end{figure}

\begin{figure}[t]
\centerline{\psfig{figure=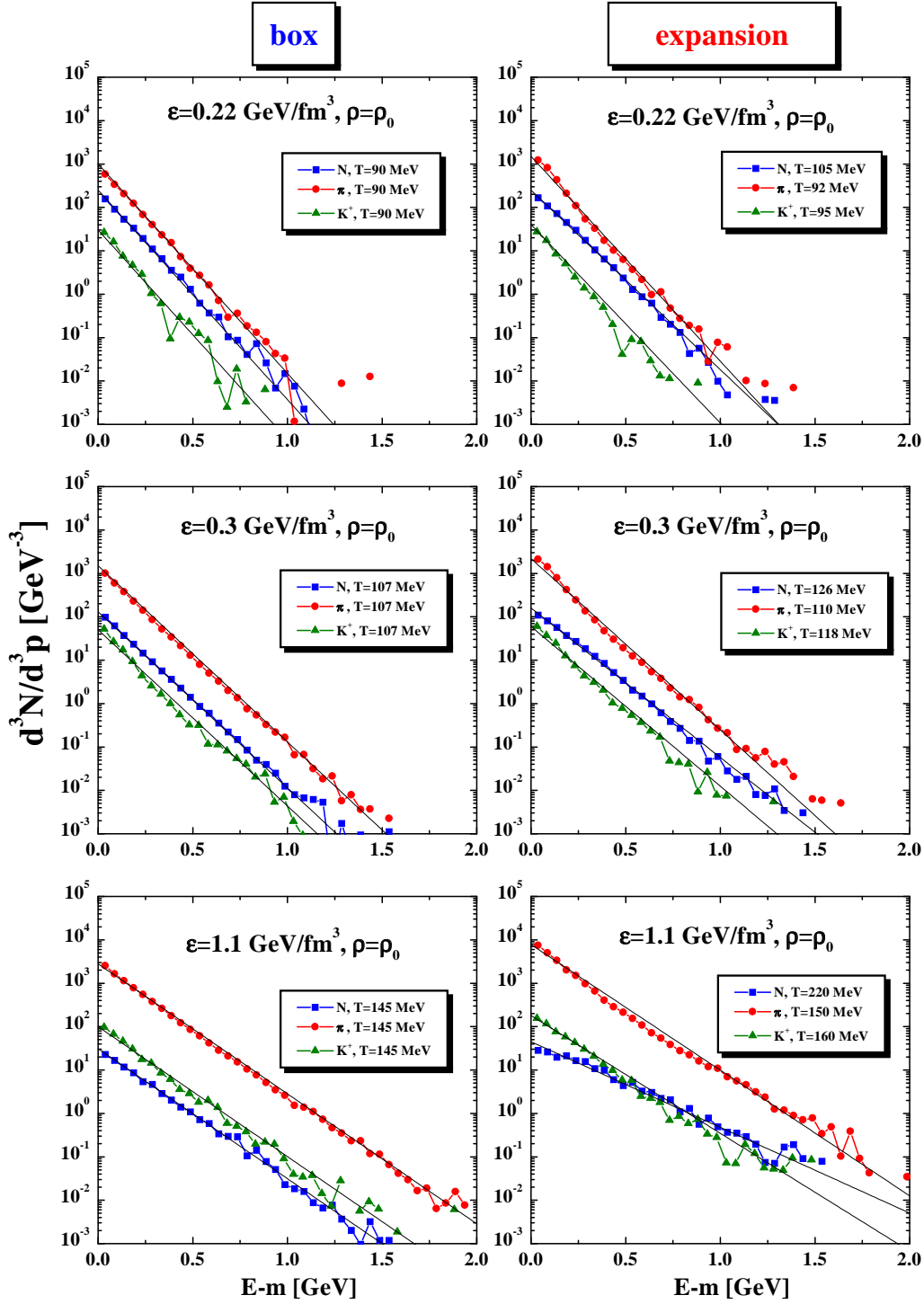,width=16cm}}
\vspace*{-2.5cm}
\caption{The spectra of nucleons ($N$), pions ($\pi$) and kaons ($K^+$)
versus the kinetic energy $E-m$ for $\rho=\rho_0$ at
$\varepsilon=0.22, 0.3$ and 1.1~GeV/fm$^3$  before the expansion
(left panel) and after the expansion (right panel).}
\label{Fig12}
\end{figure}

\begin{figure}[t]
\centerline{\psfig{figure=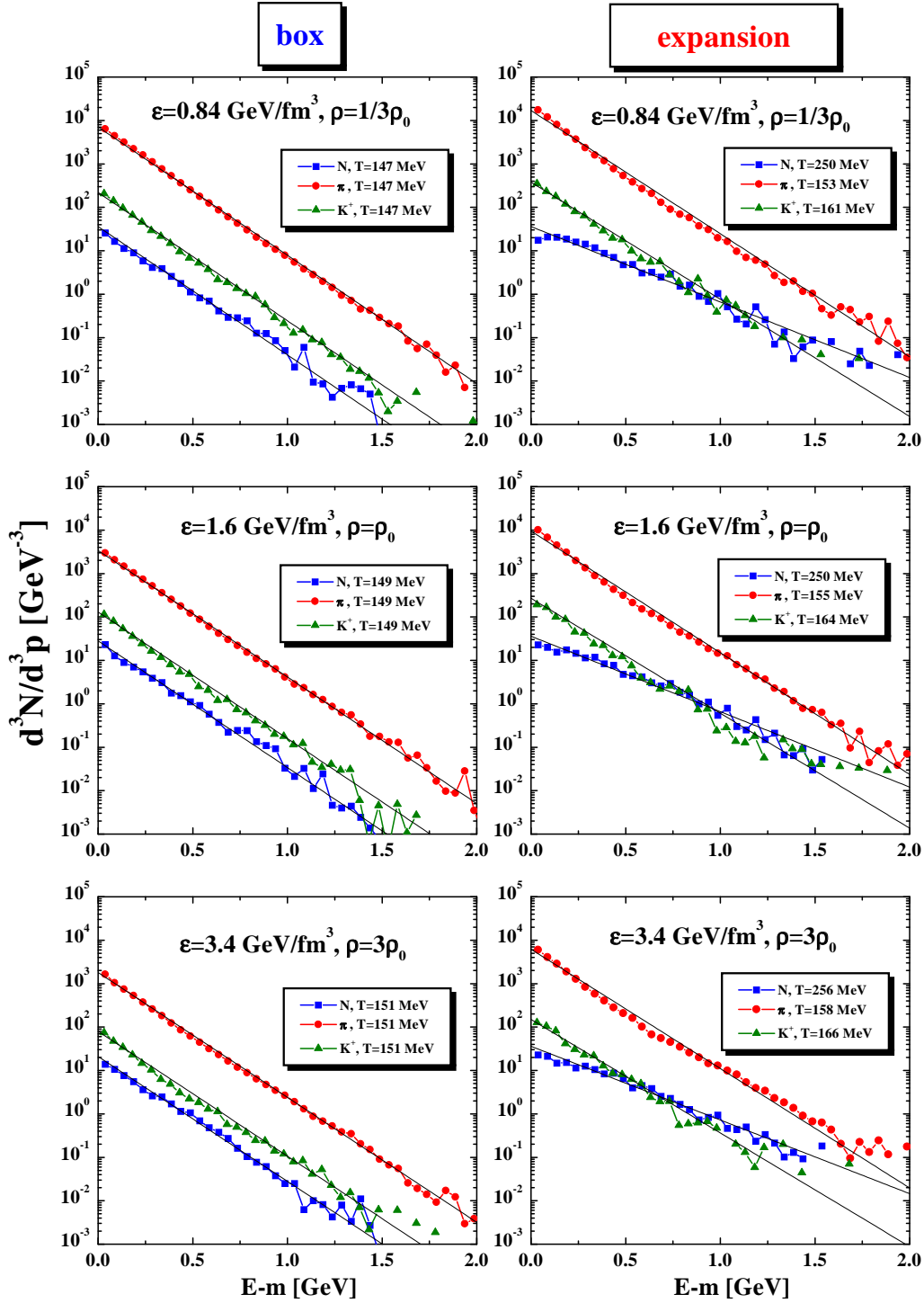,width=16cm}}
\vspace*{-2.5cm}
\caption{The spectra of nucleons ($N$), pions ($\pi$) and kaons ($K^+$)
versus the kinetic energy $E-m$ before (left panel) and after (right
panel) expansion for $\rho=1/3\rho_0$ at $\varepsilon=0.84$ GeV/fm$^3$
(upper part), for $\rho=\rho_0$ at $\varepsilon=1.6$ GeV/fm$^3$ (middel
part) and for $\rho=3\rho_0$ at $\varepsilon=3.4$ GeV/fm$^3$ (lower part). }
\label{Fig13}
\end{figure}

\begin{figure}[t]
\centerline{\psfig{figure=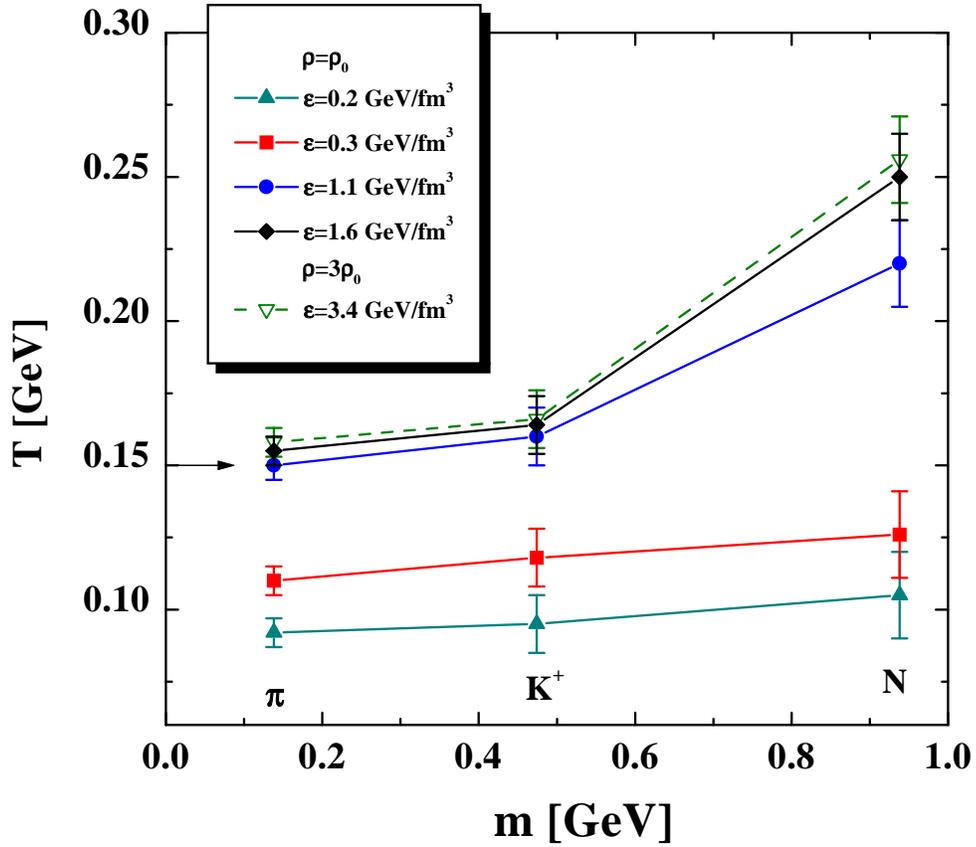,width=16cm}}
\vspace*{-2.5cm}
\caption{The spectral slope $T$ after expansion versus the hadron mass $m$
for $\pi, K^+, N$ at $\rho=\rho_0$ and different
energy densities: $\varepsilon = 0.2$~GeV/fm$^3$ (up full triangles),
$\varepsilon = 0.3$~GeV/fm$^3$ (full squares),
$\varepsilon = 1.1$~GeV/fm$^3$ (full dots),
$\varepsilon = 1.6$~GeV/fm$^3$ (full diamonds);
for $\rho=3\rho_0$ at $\varepsilon=3.4$~GeV/fm$^3$ (down open triangles).
The arrow indicates the limiting temperature $T_s\simeq 150$~MeV
before the expansion.}
\label{Fig14}
\end{figure}

\begin{figure}[t]
\phantom{a}\vspace*{-3cm}
\centerline{\psfig{figure=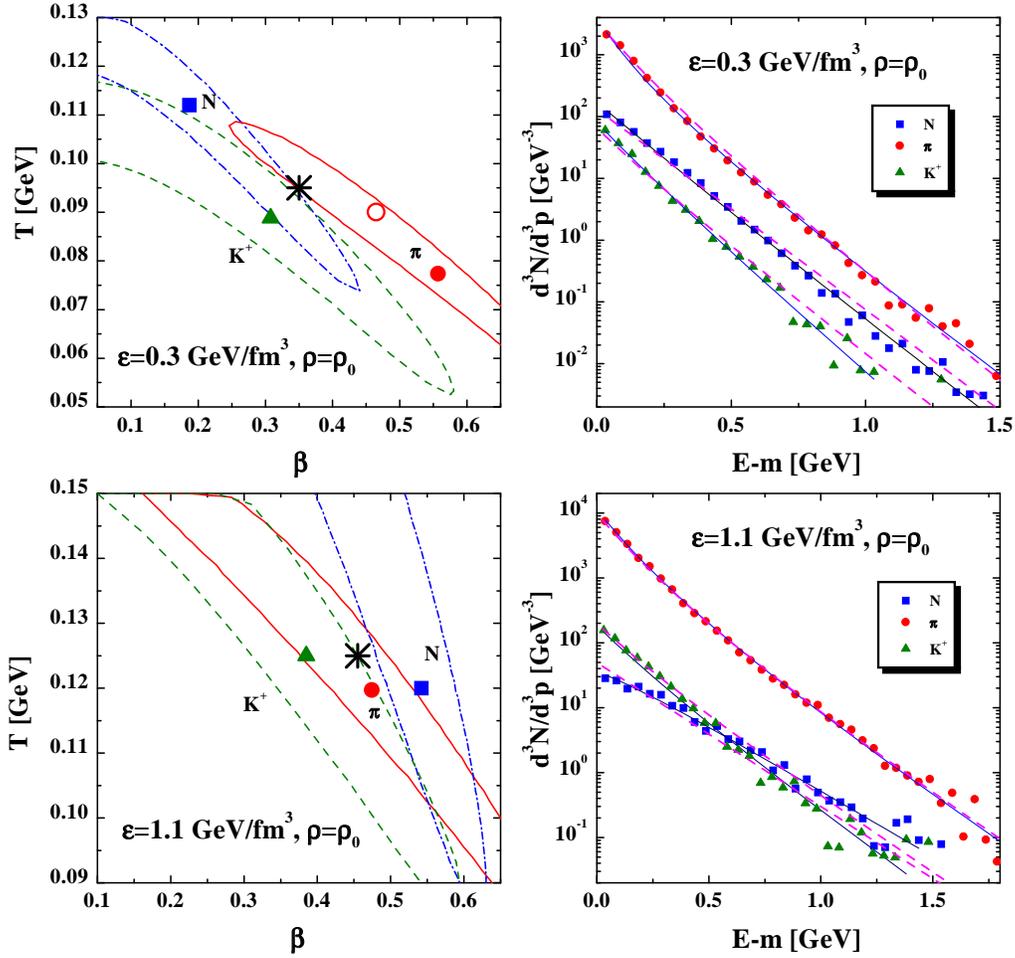,width=16cm}}
\vspace*{-4cm}
\caption{Left panel: the contour plots for the parameter errors in the
$T-\beta$ plane; dot-dashed lines: nucleons ($N$), solid lines:
pions ($\pi$) and dashed lines: kaons ($K^+$) for the energy
densities $\varepsilon =0.3$ (upper part) and 1.1~GeV/fm$^3$ (lower
part) at $\rho=\rho_0$. The full symbols indicate the 'optimal'
parameters $T$ and $\beta$ (squares for $N$, dots for $\pi$ and
triangles for $K^+$).
The open dot (upper left plot) reflects $\beta,T$ for the pion spectra
including the cut-off $E-m >0.4$ GeV.
Right panel: the full symbols (squares
for $N$, dots for $\pi$ and triangles for $K^+$) are the box
calculations (for the same $\rho$ and $\varepsilon$ as in the left
panel).  The thin solid lines show the fit of the particle spectra with the
'optimal' $T$ and $\beta$ parameters from the left panel; the dashed lines
correspond to the 'eye' fit with the average $\beta$ and $T$ parameters
given by the stars from the left panel.}
\label{Fig15}
\end{figure}

\begin{figure}[t]
\centerline{\psfig{figure=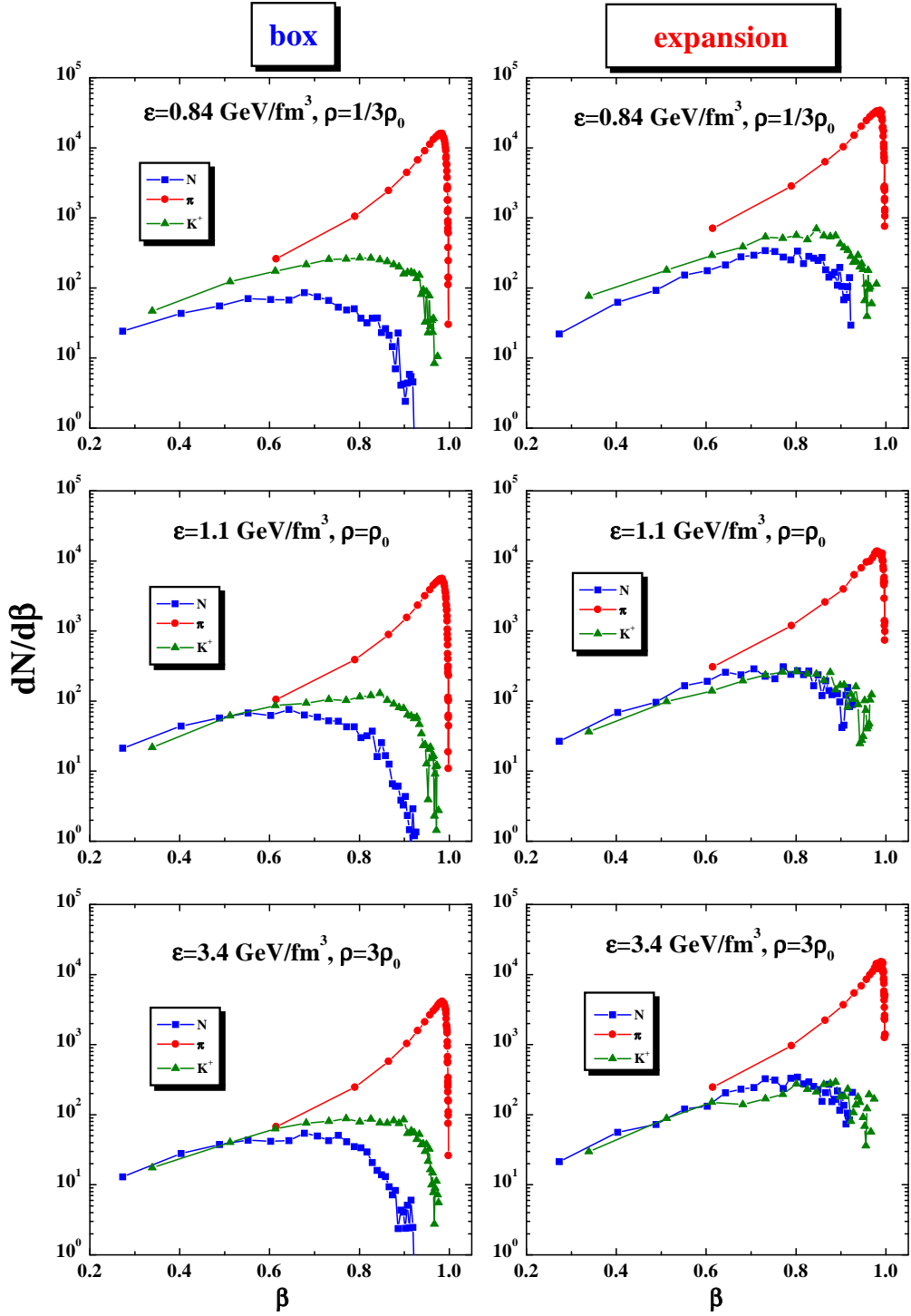,width=16cm}}
\vspace*{-2.5cm}
\caption{The velocity distributions $dN/d\beta$ for nucleons ($N$),
pions ($\pi$) and kaons ($K^+$) for $\rho=1/3 \rho_0$ at
$\varepsilon=0.84$~GeV/fm$^3$ (upper part), for $\rho=\rho_0$ at
$\varepsilon=1.1$~GeV/fm$^3$ (middle part) and for $\rho=3\rho_0$
at  $\varepsilon=3.4$~GeV/fm$^3$ (lower part). The left panel shows
$dN/d\beta$ at equilibrium whereas the right panel corresponds to
$dN/d\beta$ after the expansion phase.}
\label{Fig16}
\end{figure}

\begin{figure}[t]
\centerline{\psfig{figure=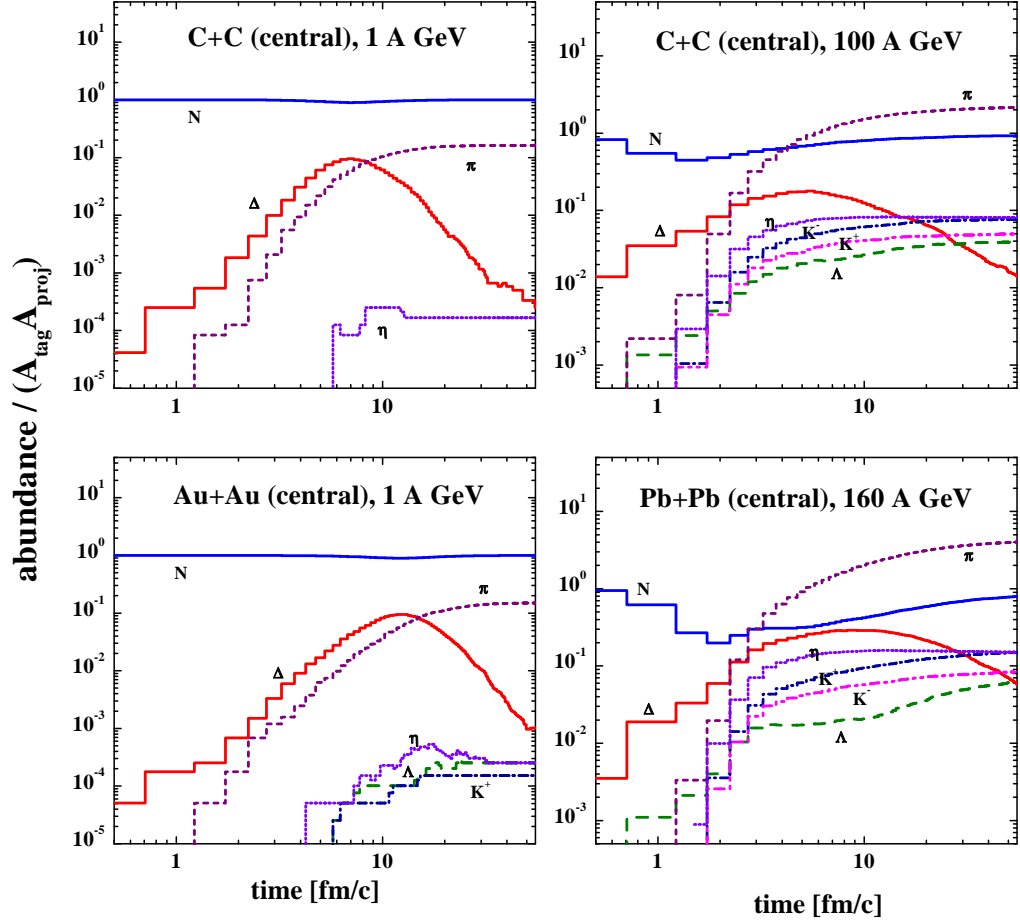,width=16cm}}
\vspace*{-2.5cm}
\caption{Time evolution of the particle abundances (nucleons $N$,
$\Delta$,  $\Lambda$, $\pi$, $\eta$,  $K^+$) for central C~+~C
collisions (upper part) and Au~+~Au and Pb~+~Pb collisions (lower part)
at 1 A$\cdot$GeV (left panel) and 160 A$\cdot$GeV (right panel).}
\label{Fig17}
\end{figure}

The thermal equilibration time in this energy range around 0.25
GeV/fm$^3$, as obtained from the box calculation, is about 5--7 fm/c
(see Fig. \ref{Fig5}). Notice, however, that this calculation --
because of its periodic boundary conditions -- probably underestimates
the equilibration time. Indeed, studies of the longitudinal and
transverse temperatures ($T_L$ and $T_T$, resp.)
\cite{Lang91,EffePhD} have shown that full thermal equilibrium is
reached only in the very late expansion phase, when the density $\rho$
has dropped already below its saturation value. After a period of 10
fm/c (after first contact) one still finds $T_L \approx 1.5\, T_T$, i.e.
an anisotropy of about 40 \%, considerably more than indicated in
Fig.~\ref{Fig5}. At the bombarding energy of 160 A$\cdot$GeV we find
a rapid decrease of the quadrupole moment in the momentum space of all
hadrons by about a factor of 3 at the scale of 5 fm/c leading to
longitudinally expanding matter. In view of Fig.~\ref{Fig5} this
stretched ellipsoid in momentum space becomes isotropic only on the
scale  of 10--20 fm/c since low energy hadronic reactions are less
effective for equilibration. Since this 'hadronic' equilibration time
is larger than the reaction time for Pb~+~Pb at 160 A$\cdot$GeV a
substantial anisotropy remains in the hadron momentum distributions
after the collision.

\begin{figure}[t]
\phantom{a}\vspace*{-2.5cm}
\centerline{\psfig{figure=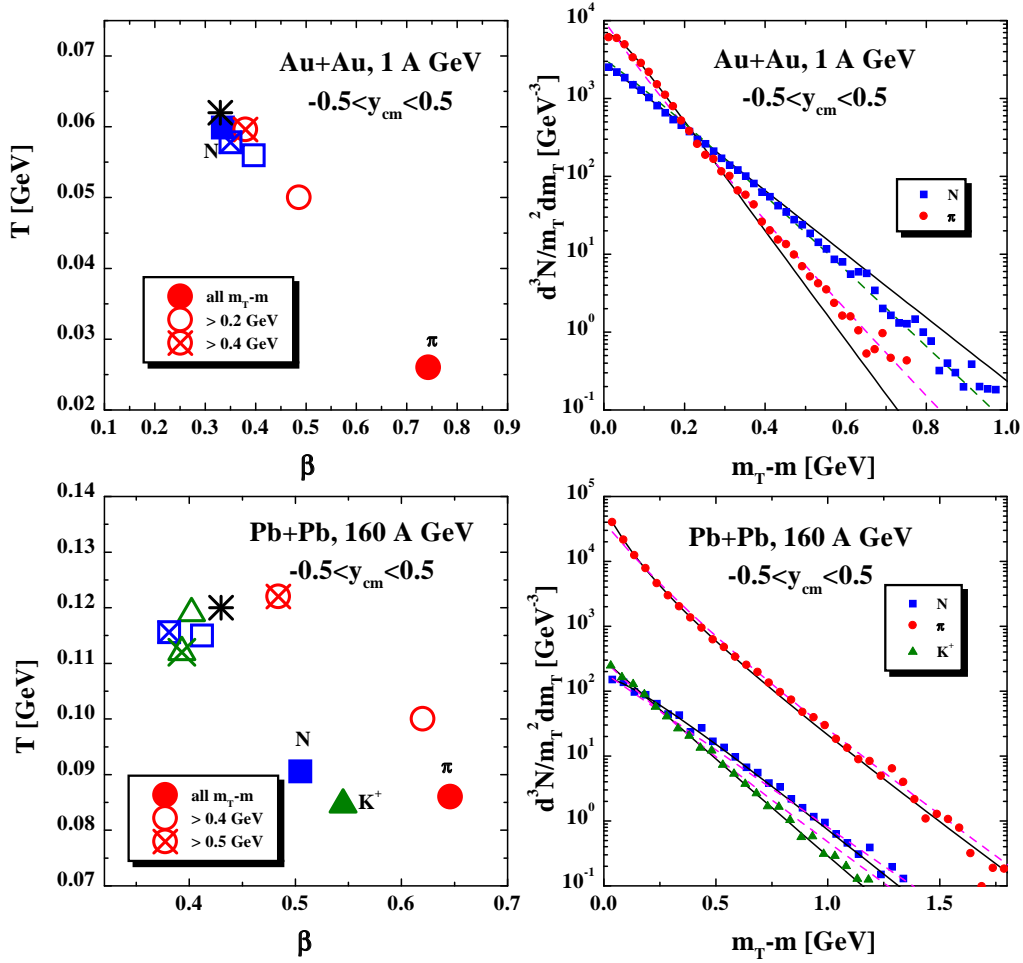,width=16cm}}
\vspace*{-6cm}
\caption{Left panel: the full symbols indicate the 'optimal' parameters
$T$ and $\beta$ (squares for $N$, dots for $\pi$ and triangles for
$K^+$) obtained by exploring the full $m_T$ spectra in the fitting
procedure. The open symbols correspond to $\beta,T$ for the cut-off $m_T-m
>0.2$~GeV (upper part) and 0.4 GeV (lower part); the open symbols with
crosses inside indicate the $\beta,T$ parameters
for the cut-off $m_T-m >0.4$~GeV
(upper part) and 0.5 GeV (lower part).  The stars corresponds to the
$\beta,T$ parameters from Ref.  \protect\cite{Averbeck} for Au~+~Au at
1 A$\cdot$GeV and from Ref. \protect\cite{Kaempfer} for Pb~+~Pb at 160
A$\cdot$GeV.  Right panel: Transverse mass ($m_T$) spectra of nucleons
($N$), pions ($\pi$) and kaons ($K^+$) for central Au~+~Au collisions
at 1 A$\cdot$GeV (upper part) and  central Pb~+~Pb collisions at
160~A$\cdot$GeV (lower part) for $-0.5\le y_{cm} \le 0.5$:  the full
symbols are the transport calculation; the thin solid lines show the
$m_T$ spectra for the 'optimal' $T$ and  $\beta$ parameters from the
left panel; the dashed lines show the fit with the $\beta,T$ values
corresponding to the 'stars' from the left panel. }
\label{Fig18}
\end{figure}

Contrary to the box case the $d^3N/dp^3$ spectra for realistic
nucleus-nucleus collisions do not follow the simple exponential
behaviour (\ref{Boltz}) due to the strong longitudinal expansion;
especially at high bombarding energy the particle spectra show the
specific 'banana' shape (reflecting the $pp$ spectra at high energies).
In order to exclude this simple dynamical effect related to the
longitudinal expansion, we present in Fig.~\ref{Fig18} (right panel)
the transverse mass spectra $1/m_T^2 dN/dm_T$ at mid-rapidity ($-0.5\le
y_{cm} \le 0.5$) versus $m_T-m$ for central Au~+~Au collisions at 1
A$\cdot$GeV (upper part) and for central Pb~+~Pb at 160 A$\cdot$GeV
(lower part) calculated at the end of the reaction. The $m_T$-spectra
show an exponential behaviour \cite{Brat_mt} (excluding small $m_T$),
however, with different slopes which can not be associated directly
with a temperature of a hot fireball formed at the intermediate stages
of the reaction.

In line with Section 3 we have performed a fit within the blast model
(\ref{flow}) within an interval of unit rapidity around midrapidity
using MINUIT.  The results of the fit are displayed in
Fig.~\ref{Fig18}. The full symbols (squares for $N$, dots for $\pi$ and
triangles for $K^+$) correspond to the 'best' values for $T$ and
$\beta$ according to the $\chi^2$ criteria. The thin solid lines in the
right panel demonstrate the fit of the $m_T$ spectra within the optimal
fit parameters (we obtain a smaller $\chi^2$ within the linear $\chi^2$
method ($\chi^2=\sum\limits_i ( f(x_i) - f_0(x_i))^2$), which provides
a better description of the low $m_T$ spectra).

Similar to Section 3 (cf. Fig.~\ref{Fig15}) we obtain (within the
'optimal' $\chi^2$ criterium) quite different freeze-out parameters for
$N, \pi$ and $K^+$ spectra. In order to exclude the influence of
$\Delta$- (and other resonance) decays on the pion spectra and to
investigate the sensitivity of the freeze-out parameters to the low
energy cuts applied, we performed a fit of the particle spectra using
the following cut-offs: $m_T-m > 0.2$~GeV (open symbols) and $m_T-m >
0.4$~GeV (open symbols with crosses inside) for Au~+~Au at 1
A$\cdot$GeV;  $m_T-m > 0.4$~GeV (open symbols) and $m_T-m
> 0.5$~GeV (open symbols with crosses inside) for Pb~+~Pb at 160
A$\cdot$GeV.  As seen from the left panel of Fig.~\ref{Fig18} the
implementation of the low $m_T$ cut-off leads to a substantial shift of
the 'optimal' MINUIT parameters $\beta$ and $T$ especially for pions.
The $\beta, T$ values for the different spectra move towards to each
other when discarding the low $m_T$ points. For Au~+~Au at 1
A$\cdot$GeV our $\beta$ and $T$ parameters agree with those extracted
by the TAPS collaboration \cite{Averbeck} using the blast model (star
in the upper left plot). Here we have to mention that the cut-off
$m_T-m > 0.4$~GeV has been applied in the experimental analysis, too
\cite{Averbeck}. For the Pb~+~Pb spectra at 160 A$\cdot$GeV our
freeze-out parameters are similar to those from K\"ampfer
\cite{Kaempfer} ($T=120$~MeV, $\beta=0.43$; star in the lower left plot).
The dashed lines in the right panel of Fig.~\ref{Fig18} show
the fit to the particle $m_T$ spectra for the $\beta,T$ values
corresponding to the 'stars' from the left panel. Again this 'eye' fit
gives a reasonable description of the spectra (except of the very low
$m_T$ part).

Here we have to mention again that the extraction of freeze-out
parameters from the experimental data is very sensitive to the
details of the thermodynamical model applied as well as to the
observables considered. For example, the analysis of SIS data at
1.0 GeV from Ref. \cite{Cleymans} gives thermal freeze-out parameters --
$T\simeq 52$ MeV and $\beta \simeq 0.4$.
At SPS energies the chemical freeze-out temperature extracted in Ref.
\cite{BM99} from the thermal-analysis of particle ratios is $T\simeq
168$ MeV, whereas the analysis of particle spectra and two-particle
correlations (HBT data) \cite{Nix,Heinz99} provides a much lower
thermal freeze-out temperature $T\simeq 90-95$~MeV. For a survey
different freeze-out parameters the reader is referred to
Fig. 4 of Ref.~\cite{Redlich}).

In view of the various uncertainties inherent in the extraction of the
thermal freeze-out parameters we conclude that a full, i.e.\ thermal
and chemical, thermodynamical equilibrium at freeze-out cannot be
deduced from such an analysis.

\section{Summary}

In this paper we have performed a systematic study of equilibration
phenomena and equilibrium properties of 'infinite' hadronic matter as
well as of relativistic nucleus-nucleus collisions using a BUU
transport model that contains resonance and string degrees-of-freedom.
The 'infinite' hadron matter is modelled  by initializing the system at
fixed baryon density, strange density and energy density by confining
it in a cubic box with periodic boundary conditions.

We have shown that the equilibration times $\tau_{eq}$ for different
particles depend on baryon density and energy density. The time
$\tau_{eq}$ for non-strange particles is much shorter than for
particles including strangeness; for kaons and antikaons the
equilibration time is found to be larger than $\simeq$ 40 fm/c for all
baryon and energy densities considered. The overall abundance of the
dominant strange particles (kaons and $\Lambda$'s) being produced and
obtained within the BUU cascade model for heavy-ion collisions can
therefore not be described by assuming a perfect chemical equilibrium
as strangeness is typically still undersaturated to a quite large
extent. We mention taht transport model calculations like ours can
describe the yield and spectra of the produced nonstrange hadrons as
well as $K^+, K^-, \Lambda$ yields quite well at SPS energies
\cite{Cass99,Geiss}.  On the other hand, at AGS energies the measured
$K^+/\pi^+$ ratio in central Au~+~Au collisions is underestimated by
about 30\% \cite{CassQM}.  However, we have to point out that the more
exotic strange particles (like the measured antihyperon yields of
Ref.~\cite{WA97}) can by far not be explained within such standard
hadronic multiple channel reactions.  These hadronic data possibly
point towards new physics.

We have, furthermore, shown that thermal equilibrium is established
quickly, within about 5 fm/c at SIS energies and samewhat larger times
at high energies.  The inclusion of continuum excitations, i.e. hadron
'strings', leads to a limiting temperature of $T_s \simeq 150$~MeV in
our transport approach which practically does not depend on the baryon
density and energy.  We have compared our results with the
statistical model (SM), which contains the same degrees of freedom and
the same spectral functions of particles as our transport model. We
found that the limiting temperature behaviour can be reproduced in the
statistical model only after including continuum excitations of the
Hagedorn type, otherwise the fireball temperature extracted from the
particle abundances and spectra is overestimated substantially.

Close to the critical temperature $T_s$, the hadronic energy densities
can increase to a couple of GeV/fm$^3$. From lattice QCD calculations
one expects that a phase transition to a potentially deconfined QGP
state should occur. Referring to the limiting temperature $T_s\approx
150 $ MeV obtained, a QGP should be revealed and clearly distinguished from
a hadronic state of matter if one can unambiguously prove the
existence of an equilibrated and thermal phase of strongly interacting
matter with temperatures exceeding, e.g., 200 MeV.  The best candidates
are electromagnetic probes, either direct photons or dileptons. On the
other hand these are also `contaminated' by hadronic background and/or
preequilibrium physics. So far no thermal electromagnetic source with
temperatures larger or equal than 200 MeV has been clearly identified.

We have also studied the expansion of the equilibrated hadronic
fireball and found that the slope parameters of the particles after
expansion increase with their mass; the pions leave the fireball much
faster than nucleons and accelerate heavier hadrons by rescattering
('pion wind'). If the system before expansion is close to the limiting
temperature $T_s$, the slope parameters for all particles after
expansion practically do not depend on energy and baryon density. This
is due to the fact that the particle velocity distributions in
equilibrium do not change any more for $T \approx T_s$. We have fitted
the resulting spectra within the blast model of Siemens and Rasmussen.
Our analysis shows a strong sensitivity of the $(\beta, T)$ parameters
on the spectral shape at low energy (or a low energy cut-off) so that
no reliable parameter determination can be reported. However, a global
'eye' fit with 'average' $(\beta, T)$ parameters describes the data
reasonably well.

Additionally, we have considered the  equilibration in realistic
nucleus-nucleus collisions of light (C~+~C) and heavy (Au~+~Au and
Pb~+~Pb) systems. The $(\beta, T)$ parameters extracted from our
calculations for Au~+~Au at 1 A$\cdot$GeV agree with those extracted from
the TAPS collaboration \cite{Averbeck} and for Pb~+~Pb at 160 A$\cdot$GeV
with the parameters from Ref.~\cite{Kaempfer}. Here the reaction time
is a few 10~fm/c and decreases with the initial energy due to the fast
expansion.  Since the reaction time is much shorter than the
equilibration time for strangeness, a chemical equilibrium of strange
particles in heavy-ion collisions is not supported by our transport
calculations. Although again simple fits within the blast model provide
a decent parametrization of our transport results for the differential
particle spectra (Fig.~\ref{Fig18}) a deduction of global parameters
for thermal freeze-out is again found to be rather ambiguous,
especially when considering also the lower momentum contributions of
the various particle spectra.

\section*{Acknowledgements}
The authors are grateful for valuable discussions with V.~Metag,
H.~Oeschler and H.~St\"ocker.


\newpage

\end{document}